\documentclass[12pt, a4paper]{article}

\usepackage[a4paper, top=2.0cm, bottom=1.5cm, left=1.5cm, right=1.5cm, footskip = 1.0cm, includehead, includefoot]{geometry}
 
\usepackage{helvet}

\usepackage{csquotes}
\usepackage[utf8]{inputenc}
\usepackage[final]{graphicx}
\usepackage[margin=20mm, font=footnotesize, labelfont=bf, justification=justified]{caption}
\usepackage{dirtytalk}

\usepackage{times}
\usepackage[bottom]{footmisc}	
\usepackage{scrextend}
\deffootnote[3.5em]{3.5em}{2em}{\textsuperscript{\thefootnotemark}}

\usepackage{amsmath, amssymb, amsthm, graphicx, color, bm, soul}
\usepackage{amsmath}
\usepackage{authblk}
\usepackage{xcolor}			
\usepackage{caption}
\usepackage{fourier}
\usepackage{mathtools} 
\usepackage{float}
\usepackage{url}
\usepackage{hyperref}

\usepackage{titlesec} 
\titleformat{\paragraph}
{\normalfont\normalsize\bfseries}{\theparagraph}{1em}{}

\usepackage{physics}
\usepackage{braket} 

\usepackage{musicography}  

\usepackage{tikz} 
\usetikzlibrary{quantikz}

\title{Teaching Qubits to Sing: Mission Impossible?}

\author{Eduardo Reck Miranda\qquad Brian N. Siegelwax\\
Interdisciplinary Centre for Computer Music Research (ICCMR), \\
Faculty of Arts, Humanities and Business,\\
University of Plymouth, Plymouth, PL4 8AA,\\
United Kingdom\\
eduardo.miranda@plymouth.ac.uk, bsiegelwax@gmail.com}

\begin{document}

\maketitle

\abstract{This paper introduces QuSing, a system that learns to sing new tunes by listening to examples. QuSing extracts sequencing rules from input music and uses these rules to generate new tunes, which are sung by a vocal synthesiser. We developed a method to represent rules for musical composition as quantum circuits. We claim that such musical rules are \textit{quantum native}: they are naturally encodable in the amplitudes of quantum states. To evaluate a rule to generate a subsequent event, the system builds the respective quantum circuit dynamically and measures it.  After a brief discussion about the vocal synthesis methods that we have been experimenting with, the paper introduces our novel generative music method through a practical example.  The paper shows some experiments and concludes with a discussion about harnessing the system's creative potential. Accompanying materials are available in an Appendix. Audio recordings of the musical examples and programming code are available: \url{https://github.com/iccmr-quantum/QuSing}.}

\section{Introduction}

\medskip
Research and development in computer music and professional usages have been progressing in tandem with Computer Science since the invention of the computer. Musicians started experimenting with computers far before the emergence of the vast majority of scientific, industrial and commercial computing applications in existence today. For instance, in the 1940s, researchers at Australia’s Council for Scientific and Industrial Research (CSIR) installed a loudspeaker on their Mk1 computer to track the progress of programs using sound. Subsequently, Geoff Hill, a mathematician with a musical background, programmed this machine to play back a tune in 1951 \cite{Doornbusch2004}.

\medskip
The first uses of computers in music were for composition. The great majority of computer music pioneers were composers interested in developing innovative approaches to composition \cite{HillerIsaacson1959, Manning2004, Mathews1970}. Nowadays, computing technology is omnipresent in almost every aspect of the music industry \cite{Bevins2013, Wikstrom2013}. Therefore, ever-evolving quantum computing technologies will continue to impact how we create, perform, listen and commercialize music in time to come. In the newborn field of \textit{Quantum Computer Music} \cite{Hamido2020, Kirke2017, Mannone2022, Miranda2021, Miranda2020a}, researchers and practitioners are exploring ways to leverage the quantum-mechanical nature of quantum computing to compose, perform, analyse and synthesise music and sound.

\medskip

Classical computers manipulate information represented in terms of binary digits, each of which can value $1$ or $0$. They work with microprocessors made up of billions of tiny switches that are activated by electric signals. Values $1$ and $0$ reflect the on and off states of the switches.

\medskip
In contrast, a quantum computer deals with information in terms of quantum bits, or qubits. Qubits operate at the subatomic level. Therefore, they are subject to the laws of quantum physics.

\medskip
At the subatomic level, a quantum object does not exist in a determined state. Its state is unknown until one observes it. Before it is observed, a quantum object is said to behave like a wave. But when it is observed it becomes a particle. This phenomenon is referred to as wave-particle duality.

\medskip
Quantum systems are described in terms of wave functions. A wave function expresses the state of a quantum system as the sum of the possible states that it may fall into when it is observed. Each possible component of a wave function, which is also a wave, is scaled by a coefficient reflecting its relative weight. That is, some states might be more likely than others. Metaphorically, think of a quantum system as the spectrum of a musical sound, where the different amplitudes of its various wave components give its unique timbre. As with sound waves, quantum wave components interfere with one another, constructively and destructively. In quantum mechanics, the interfering waves are said to be coherent. The act of observing waves decoheres them. Again metaphorically, it is as if when listening to a musical sound one would perceive only a single spectral component; probably the one with the highest energy, but not necessarily so.

\medskip
Qubits are special because of the wave-particle duality. Qubits can be in an indeterminate state, represented by a wave function until they are read out. This is known as superposition. A good part of the art of programming a quantum computer involves manipulating qubits to perform operations while they are in such an indeterminate state. This makes quantum computing fundamentally different from digital computing.

\medskip
An introduction to the nuts and bolts of quantum computing is beyond the scope of this paper. This can be found in \cite{Bernhardt2019, Grumbling2019, Rieffel2011, Siegelwax2021, Wong2022}.

\medskip
This paper presents QuSing, a system that learns to sing tunes from given examples. The system extracts compositional rules from given pieces of music and uses these rules to generate new tunes rendered with a vocal synthesiser (Figure \ref{fig:overal_flowchart}). The new vocal tunes resemble the original music and the system provides the means to vary the degree of resemblance. In a nutshell, the compositional rules are extracted classically, so to speak, but the engine that processes the rules to generate the new tunes are processed quantumly.

\medskip
A preliminary attempt at using quantum computing to control a vocal synthesiser was presented in \cite{Miranda2021}. In that case, the parameters for vocal synthesis were derived from bitstrings produced using a quantum 'hyper-dice' consisting of nine qubits. Albeit simplistic, this work paved the way for the development of QuSing.

\medskip
The paper begins by presenting the vocal synthesisers that are used to render the singing, followed by reviewing the notion of musical composition with transition rules.  Then, it introduces the quantum computing method that we invented to generate the new tunes. Next, it walks through a detailed example illustrating how the system works. The paper concludes with a discussion and some final remarks, which includes work that is in progress. The Appendices contain accompanying materials. Audio recordings of the musical examples are available on SoundClick \cite{MusicExamples} and programming code can be found from QuSing's GitHub repository \cite{QuSingQuTune}.

\section{Vocal Synthesis}

QuSing has two options for synthesising the singing. One uses a bespoke implementation of a well-known formant synthesis method, often referred to as Klatt synthesis \cite{Klatt1980}. The other uses a commercially available system called Vocaloid \cite{Kenmochi2007}, which is based on a method known as concatenative synthesis \cite{Schwarz2005}\footnote{The accompanying sound examples for this paper were produced using the latter.}. 

\begin{figure}[H]
\begin{center}\vspace{0.8cm}
\includegraphics[width=0.4\linewidth]{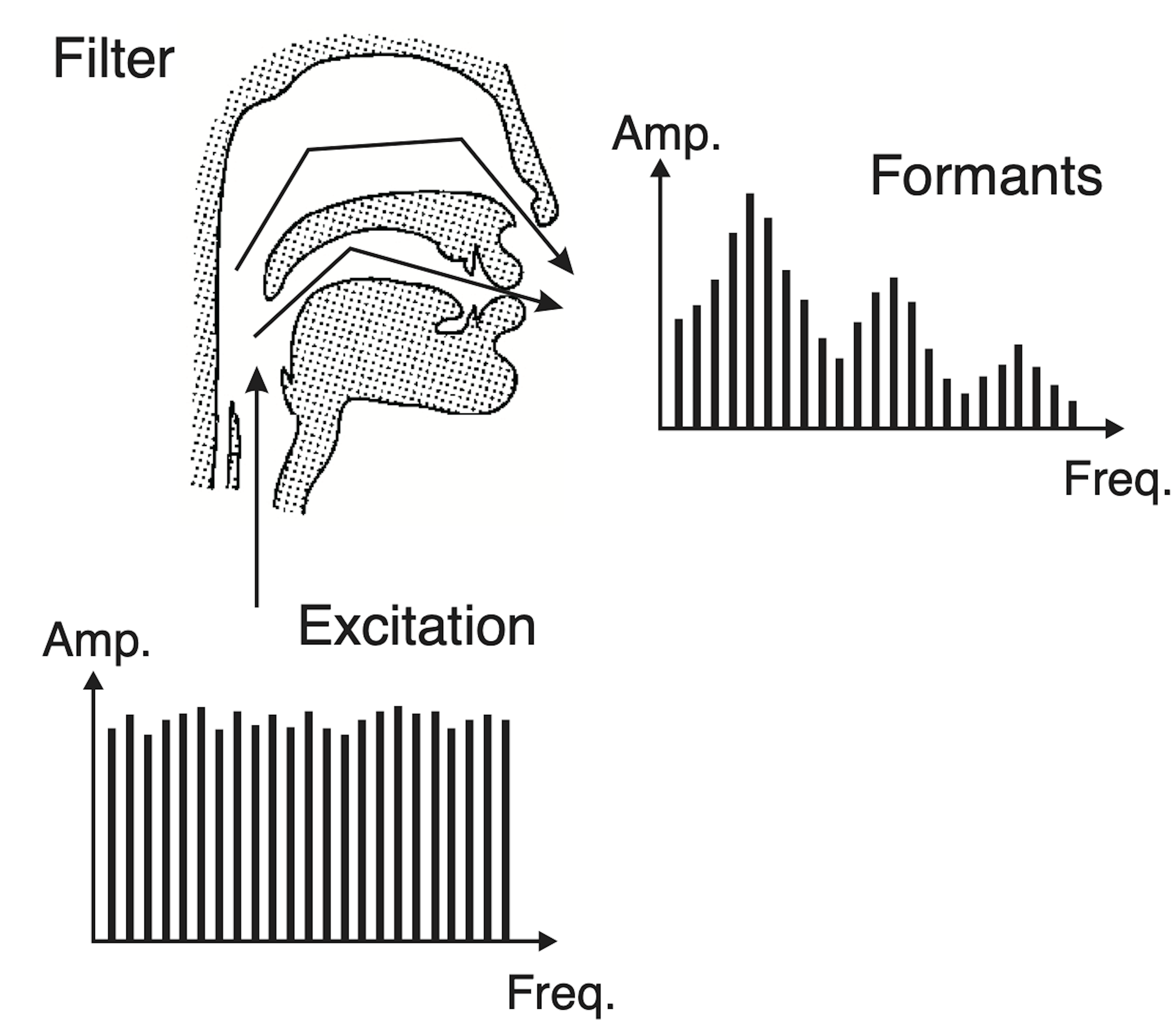}
\caption{When singing, excitation is forced upwards through the trachea from the lungs. On its journey through the vocal tract, the excitation signal is transformed, producing formants.}
\label{fig:vocal-tract}
\end{center}
\end{figure}

\medskip
The frequency spectrum of a vocal sound has the appearance of a pattern of ‘hills and valleys’, technically called \textit{formants} (Figure \ref{fig:vocal-tract}). When speaking or singing, air streams are forced upwards through the trachea from the lungs. Technically, such streams are comparable to audio signals and are referred to as \textit{excitation signals}.

\medskip

At the base of the larynx, the vocal cords are folded inwards from each side, leaving a variable tension and a slit-like separation, controlled by muscles. In normal breathing, the cords are held apart to permit the free flow of air. In singing or speaking, the cords are brought close together and tensed. The forcing of the excitation through the vocal cords in this state sets them into vibration. As a result, the excitation signals are modulated at the vibration frequency of the vocal cords. This motion determines the pitch of a vocal sound.

\medskip

The vocal system can be thought of as a resonating structure in the form of a pipe (Figure \ref{fig:vocal-tract}), from the vocal cords to the lips, plus a side branch leading to the nose. On its journey through the vocal tract, the excitation is transformed (that is, filtered) by the resonance of the vocal `pipe'. Components that are close to the resonance frequencies of the tract are transmitted with high amplitude, while those which lie far from the resonance frequencies are suppressed, hence the hills and valleys pattern in the frequency spectra of vocal sounds. The art of singing lies in controlling the energy of the excitation, the tension of the vocal folds and the shape of the vocal tract to produce the desired tones.

\medskip
Traditionally, a Klatt formant synthesiser is a \textit{subtractive synthesiser} \cite{Hahn2020} consisting of an excitation module and a resonator module. For more on Klatt's method see also \cite{Anumanchipalli2010, Klatt1980, Rutledge1995}. QuSing's implementation of Klatt's method deploys generators of excitation signals (e.g., band-limited noise and pulse generators) and a bank of band-pass filters tunable to produce formants. Originally, this method was developed for producing speech. We adapted it for synthesising singing voice.

\medskip

Alternatively, some systems focus on replicating the actual sounds rather than modelling the vocal system.  Normally, such systems hold a database of recorded basic building blocks (e.g., phonemes, onomatopoeias, vowels), which are picked and concatenated to form words, tunes, and so on. Some of them deploy a technique referred to as analysis-resynthesis \cite{Serra1990}. Think of this as something like applying a Fourier Transform \cite{FFT} to analyse the sound and then doing an inverse Fourier Transform to re-synthesise it. The system analyses those building blocks to extract spectral features and store them. Then, rather than necessarily concatenating the original sounds per se, the system concatenates the respective analysis information, which is then used to synthesise the utterance. In this case, the system can manipulate the analysed information to achieve particular effects, such as changing the prosody of the utterance, pitch, vocal timbre, and more \cite{Maestre2009, Villavicencio2010}. Essentially, the Vocaloid system is a \textit{concatenative synthesiser}, which uses analysis-resynthesis and machine-learning to process vocal fragments extracted from recordings of humans singing \cite{Wilson2017}.

\medskip

In both cases, the decoded musical events (Figure \ref{fig:overal_flowchart}) are translated onto control parameters for the respective synthesisers. In the simplest case scenarios, which are the ones reported in this paper, they control the pitches and lengths of the excitation signals (Klatt synthesiser) or the pitches and lengths of the concatenated segments (Vocaloid).

%
%
\section{Composing with Transition Rules}
\label{sec:tran_rules}

Humans possess an irresistible urge to produce and appreciate sound arrangements beyond the mere purposes of signalling or linguistic communication. It seems that we do this for no specific purpose other than to enjoy it. This is an intriguing evolutionary trait that differentiates us from other animals \cite{Brown1999}. But what is music?

\medskip
As a working definition, let us establish that \textit{music is sounds organised in space and time}. When we listen to music, our brain has certain expectations; for instance, it expects some order in the auditory stimuli. In music theory, this is generally referred to as \textit{musical form}.

\medskip
Psychology teaches us that our auditory system employs mental schemes to make sense of streams of sounds \cite{McAdams1987}. However, there is no agreement about which of such mental schemes are genetically hard-wired in our brains and which ones evolve culturally as we grow up. This suggests that there is no good or bad music; it depends on culture, individual taste, and so on. But we should not need to worry about this debate here. What is important is to concur that music creators organise sounds according to some criteria. Thus, for computers to create music one needs to endow them with such criteria; to glance over different approaches for doing this, please refer to \cite{Brown_etal_2015, DodgeJerse1997, Miranda2001, Nierhaus2009}. One such approach is to program the computer with (a) rules for sequencing musical events and (b) the ability to use those rules.

\medskip 
As an example, let us consider a given set of eight musical pitches as follows: \[\{C_3, D_3, E_3, F\musSharp{}_3, G\musSharp{}_3, A\musSharp{}_3, C_4, D_4\}\]

\medskip 

To create musical forms with those pitches, we need some rules governing how they can be sequenced; e.g., some pitches might not be allowed to follow a certain pitch, or some might have priority to follow some others, and so on. Let us define a few transition rules, as follows:

\begin{itemize}
\item $C_3 \Longrightarrow D_3(25\%) \lor G\musSharp{}_3(25\%) \lor C_4(25\%) \lor D_4(25\%)$
\item $D_3 \Longrightarrow C_3(30\%) \lor E_3(70\%)$
\item $E_3 \Longrightarrow D_3(25\%) \lor F\musSharp{}_3(25\%) \lor A\musSharp{}_3(25\%) \lor C_4(5\%) \lor D_4(20\%)$
\item $F\musSharp{}_3 \Longrightarrow E_3(100\%)$
\item $G\musSharp{}_3 \Longrightarrow C_3(30\%) \lor A\musSharp{}_3(70\%)$
\item $A\musSharp{}_3 \Longrightarrow E_3(33\%) \lor G\musSharp{}_3(33\%) \lor C_4(34\%)$ 
\item $C_4 \Longrightarrow C_3(30\%) \lor A\musSharp{}_3(70\%)$
\item $D_4 \Longrightarrow C_3(20\%) \lor E_3(80\%)$
\end{itemize}

\medskip 

The symbol \enquote{$\lor$} stands for \enquote{or} and the percentage figure in parenthesis next to the notes is their weight coefficient, expressed here in terms of probability of occurrence. For instance, the second rule states that given a pitch $D_3$, only $C_3$ or $E_3$ can follow it, but $E_3$ has a higher priority to occur. In other words, the rule is saying that given a pitch $D_3$, there is a 30\% chance that it would be followed by a $C_3$ and a 70\% chance that it would be followed by an $E_3$. For didactic purposes, with simple rule systems like the example above, it is often useful to visualise it as a \textit{transition table} (Figure \ref{table:transitionmatrix}).

\medskip 

\begin{table}[H]
\centering
\begin{tabular}{|l|l|l|l|l|l|l|l|l|l|l|l|l|}
\hline
\multicolumn{1}{|c|}{} & \multicolumn{1}{c|}{$C_3$} & \multicolumn{1}{c|}{$D_3$} & \multicolumn{1}{c|}{$E_3$} & \multicolumn{1}{c|}{$F\musSharp{}_3$} & \multicolumn{1}{c|}{\textbf{$G\musSharp{}_3$}} & \multicolumn{1}{c|}{$A\musSharp{}_3$} & \multicolumn{1}{c|}{$C_4$} & \multicolumn{1}{c|}{$D_4$} \\ \hline
\multicolumn{1}{|c|}{$C_3$} & \multicolumn{1}{c|}{} & \multicolumn{1}{c|}{$25\%$} & \multicolumn{1}{c|}{} & \multicolumn{1}{c|}{} & \multicolumn{1}{c|}{$25\%$} & \multicolumn{1}{c|}{} & \multicolumn{1}{c|}{$25\%$} & \multicolumn{1}{c|}{$25\%$} \\ \hline
\multicolumn{1}{|c|}{$D_3$} & \multicolumn{1}{c|}{$30\%$} & \multicolumn{1}{c|}{} & \multicolumn{1}{c|}{$70\%$} & \multicolumn{1}{c|}{} & \multicolumn{1}{c|}{} & \multicolumn{1}{c|}{} & \multicolumn{1}{c|}{} & \multicolumn{1}{c|}{} \\ \hline
\multicolumn{1}{|c|}{$E_3$} & \multicolumn{1}{c|}{} & \multicolumn{1}{c|}{$25\%$} & \multicolumn{1}{c|}{} & \multicolumn{1}{c|}{$25\%$} & \multicolumn{1}{c|}{} & \multicolumn{1}{c|}{$25\%$} & \multicolumn{1}{c|}{$5\%$} & \multicolumn{1}{c|}{$20\%$} \\ \hline
\multicolumn{1}{|c|}{$F\musSharp{}_3$} & \multicolumn{1}{c|}{} & \multicolumn{1}{c|}{} & \multicolumn{1}{c|}{$100\%$} & \multicolumn{1}{c|}{} & \multicolumn{1}{c|}{} & \multicolumn{1}{c|}{} & \multicolumn{1}{c|}{} & \multicolumn{1}{c|}{} \\ \hline
\multicolumn{1}{|c|}{$G\musSharp{}_3$} & \multicolumn{1}{c|}{$30\%$}  & \multicolumn{1}{c|}{} & \multicolumn{1}{c|}{} & \multicolumn{1}{c|}{} & \multicolumn{1}{c|}{}  & \multicolumn{1}{c|}{$70\%$} & \multicolumn{1}{c|}{}  & \multicolumn{1}{c|}{} \\ \hline
\multicolumn{1}{|c|}{$A\musSharp{}_3$} & \multicolumn{1}{c|}{}  & \multicolumn{1}{c|}{} & \multicolumn{1}{c|}{$33\%$}  & \multicolumn{1}{c|}{}  & \multicolumn{1}{c|}{$33\%$}  & \multicolumn{1}{c|}{} & \multicolumn{1}{c|}{$34\%$}& \multicolumn{1}{c|}{} \\ \hline
\multicolumn{1}{|c|}{$C_4$} & \multicolumn{1}{c|}{$30\%$} & \multicolumn{1}{c|}{} & \multicolumn{1}{c|}{} & \multicolumn{1}{c|}{} & \multicolumn{1}{c|}{} & \multicolumn{1}{c|}{$70\%$} & \multicolumn{1}{c|}{} & \multicolumn{1}{c|}{} \\ \hline
\multicolumn{1}{|c|}{$D_4$} & \multicolumn{1}{c|}{$20\%$} & \multicolumn{1}{c|}{} & \multicolumn{1}{c|}{$80\%$} & \multicolumn{1}{c|}{} & \multicolumn{1}{c|}{} & \multicolumn{1}{c|}{} & \multicolumn{1}{c|}{} & \multicolumn{1}{c|}{} \\ \hline
\end{tabular}\caption{\footnotesize{Transition table representing the rules.}}
\label{table:transitionmatrix}
\end{table}

\medskip
It should be said that such rules do not necessarily need to consider just one previous or just one subsequent event. There could be rules like: $\{D_3 \rightarrow C_3\} \Longrightarrow E_3(30\%) \lor G\musSharp{}_3(70\%)$, or $\{D_3 \rightarrow C_3 \rightarrow D_3\} \Longrightarrow \{E_3 \rightarrow C_4\}(70\%) \lor G\musSharp{}_3(30\%)$, and so on. 
 
\medskip
Given a set of rules, it is crucial to devise a method to compute them to generate sequences. There are many ways of doing this with `classical' computing methods, which will not be discussed here \cite{Bell2011, Hadimlioglu2018, Shapiro2021}. Rather, we are interested in exploring ways of doing it quantumly. In section \ref{sec:quantum_gen} we introduce the method that we invented for this and in section \ref{sec:system} we walk through an illustrative example. We argue and demonstrate that the rules for musical composition, of the type presented here, are \textit{quantum native}. They are naturally encodable in the amplitudes of quantum states.

%
%

\section{Computing Transition Rules Quantumly}
\label{sec:quantum_gen}

\medskip
There have been some previous attempts at designing quantum algorithms to generate music with transition rules. 

\medskip
Weaver proposed a system whereby a $4 \times 4$ matrix represents transitions rules (similar to those shown in section  \ref{sec:tran_rules}) for a set of four notes. He designed a quantum circuit using two qubits and rotation operator gates. The probabilities in the matrix are converted into angles for the rotation gates \cite{Weaver2018}. The caveat of this system is that it is limited to a set of four notes. But to a certain extent, the method that we propose below builds upon Weaver's idea. Our method scales up to any number of notes.

\medskip
Another method is the \textit{Basak-Miranda algorithm} \cite{Miranda2022}, which leverages a property of quantum mechanics known as constructive and destructive interference to compute the rules. It is based on the well-known Grover's algorithm [10], which has become a favoured example to demonstrate the advantage of quantum computing for searching for information in databases. However, the Basak-Miranda algorithm is limited to using equal weight distribution between the possible next notes. For instance, two possible notes would have a $50\% \times 50\%$ distribution between them by default.

\medskip
A different approach uses quantum annealing, which is an alternative model of quantum computation \cite{Arya2022, Chuharski2022}. Quantum annealing is suitable for propositional (Boolean) satisfiability problems\footnote{A propositional satisfiability problem, often called SAT, is the problem of determining whether a set of logic sentences is satisfiable.} and other combinatorial searching problems. The selection of suitable paths in a transition rule can be modelled as a satisfiability problem. This approach works well but requires special hardware, which jeopardises its compatibility for integration with potentially more generic gate-based quantum computing systems. A comparison between gate-based and quantum annealing models of computation is beyond the scope of this paper; please refer to \cite{Crosson2021, Grant2020, Strategist2020}.

\medskip
Let us introduce our method through an example using the same pitch set and rules used in section \ref{sec:tran_rules}. 

\medskip
Each time the system needs to evaluate a transition rule it builds and measures a quantum circuit that is specific to the rule in question. The circuit is built in such a way that the wave function that defines the state of the quantum system represents the probability distribution of the rule in question. That is, the measurement is likely to collapse the qubits to one of the allowed options for the next event, considering the probability weights. 

\medskip
First of all, we assign binary codes to the pitches, as follows:
 
 \begin{itemize}
\item $C_3 \Longrightarrow 000$
\item $D_3 \Longrightarrow 001$
\item $E_3 \Longrightarrow 010$
\item $F\musSharp{}_3 \Longrightarrow 011$
\item $G\musSharp{}_3 \Longrightarrow 100$
\item $A\musSharp{}_3 \Longrightarrow 101$ 
\item $C_4 \Longrightarrow 110$
\item $D_4 \Longrightarrow 111$
\end{itemize}

\begin{figure}[H]
\begin{center}\vspace{0.3cm}
    \begin{tikzpicture}
        \node[scale=0.9] 
        {
            \begin{quantikz}
                \lstick{$q_0$} & \ket{0} & \gate{\textbf{RY($\pi/4$)}}   & \targ{}  &  \gate{\textbf{RY($\pi/4$)}}  & \targ{}  & \gate{\textbf{RY($-\pi/4$)}}   & \qw & ... & \\ 
                \lstick{$q_1$} & \ket{0} & \qw  & \ctrl{-1} & \qw  & \qw & \qw & \qw & ... & \\ 
                \lstick{$q_2$} & \ket{0} & \gate{\textbf{RY($1.98$)}}  & \qw &  \qw & \ctrl{-2} &  \qw & \qw & ... & \\
                \lstick{}
                \lstick{\\}
            \end{quantikz}
        };  
    \end{tikzpicture}

\begin{tikzpicture}
        \node[scale=0.9] 
        {
            \begin{quantikz}
                \lstick{} & ... & &   \targ{}   & \gate{\textbf{RY($-\pi/4$)}}   & \targ{}    &  \meter{} & \qw  \\
                \lstick{} & ... & &  \ctrl{-1}  & \qw  & \qw & \meter{} & \qw \\
                \lstick{} & ... & &  \qw & \qw   & \ctrl{-2}  & \meter{} & \qw
            \end{quantikz}
    };  
    \end{tikzpicture}    
\end{center}
\caption{Circuit for the rule $100\Longrightarrow 000(30\%) \lor 101(70\%)$.}
\label{fig:example_circuit}
\end{figure}
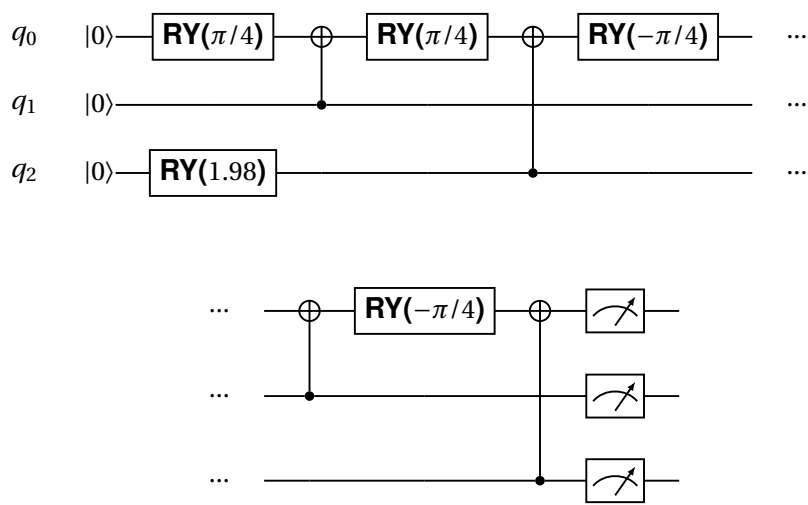

\medskip
Now, let us consider the rule $C_4 \Longrightarrow C_3(30\%) \lor A\musSharp{}_3(70\%)$. In terms of the binary representation, this rule is expressed as $110\Longrightarrow 000(30\%) \lor 101(70\%)$. In this case, the system builds the circuit shown in Figure \ref{fig:example_circuit}. As the set has eight pitches, the circuit needs only three qubits to represent every possible solution. 

\medskip
Upon measurement, the circuit (Figure \ref{fig:example_circuit}) should output either $000$ with $30\%$ probability or $101$ with $70\%$. (The ordering of the qubits are $q_2 q_1 q_0$.)

\section{The QuSing System}
\label{sec:system}

\begin{figure}[H]
\begin{center}\vspace{1.0cm}
\includegraphics[width=0.66\linewidth]{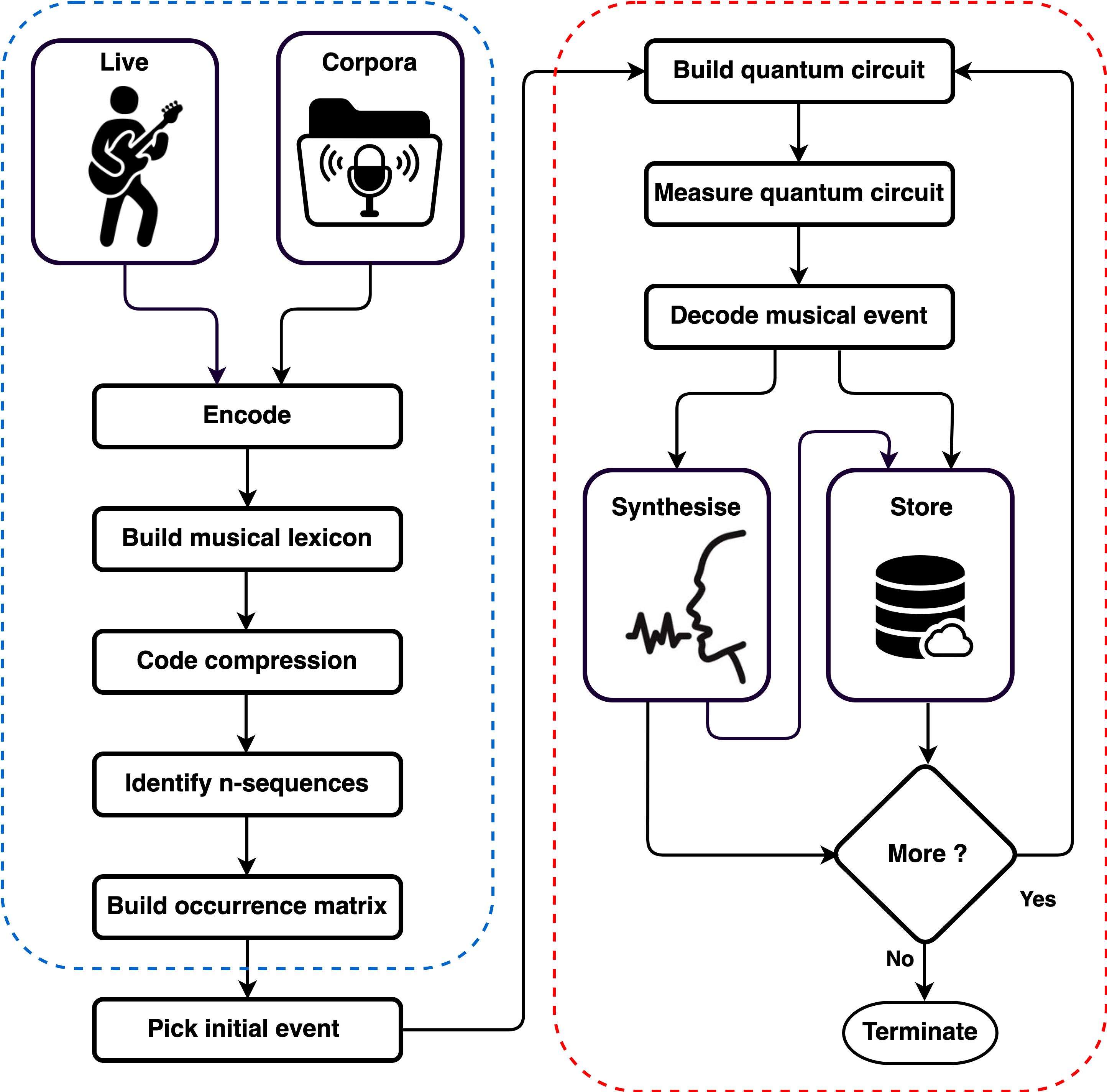}
\caption{System's flow diagram.}
\label{fig:overal_flowchart}
\end{center}
\end{figure}

A flow diagram depicting a bird's eye view of QuSing is shown in Figure \ref{fig:overal_flowchart}.  This section walks through each stage using a simple example. More examples are given in the Appendix. The quantum computing components of QuSing were implemented in Qiskit\footnote{Qiskit is an open-source software development kit (SDK) for programing quantum computers using Python.} and we ran the experiments and demonstrations discussed in this paper using IBM Quantum's resources \cite{IBMQuantum}.

\medskip

The system has two main phases, highlighted within blue and red dashed boxes: a machine learning phase (blue) and a generative one (red). In a nutshell, in the machine learning phase, the system `listens' to one or more musical tunes and extracts probabilistic rules governing their structure. Then, in the generative phase, it uses these rules to generate new tunes and `sings' them. Once the system has learned the rules, it can generate as many new songs, of virtually any length, as required. The learning is done classically and the generation of new music is done quantumly. 

\subsection{Music input}

QuSing accepts two types of inputs: live music input and a MIDI\footnote{MIDI is an acronym that stands for Musical Instrument Digital Interface. It is a technical standard that describes a communications protocol, to connect electronic musical instruments, computers, and related audio devices for playing, editing, and recording music.} music file (or a set thereof). The live input can be from a MIDI instrument or audio, through a microphone or pickup. If audio is used, then it needs to be converted to MIDI for the Encode stage\footnote{It would be possible to process audio in the Encode stage. But this would require significant effort. It is convenient to convert audio to MIDI first, as there are audio-to-MIDI technologies readily available; e.g., \cite{Bittner2022}.}

\medskip

\begin{figure}[H]
\begin{center}\vspace{0.8cm}
\includegraphics[width=0.7\linewidth]{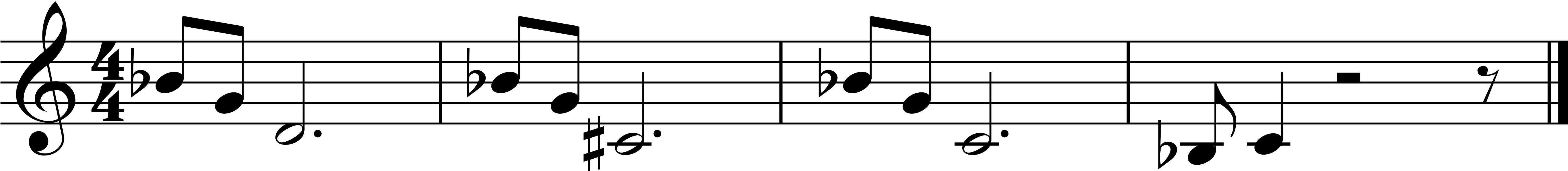}
\caption{An extract from the \textit{Mission: Impossible}'s theme.}
\label{fig:mission_input}
\end{center}
\end{figure}

The system only accepts monophonic music; i.e., single-note tunes rather than more than one note sounding simultaneous, or chords. As an example, let us input the MIDI file of an extract of the \textit{Mission: Impossible}\footnote{This is an espionage television series aired on USA TV in the 1960s and 1970s. Several films followed casting Hollywood stars such as Tom Cruise and Henry Cavill.} theme, shown in Figure \ref{fig:mission_input}.

\subsection{Encode}

Firstly, the system needs to convert the input's MIDI codes to the system's own bespoke representation, which is more efficient than MIDI for our purposes. 

\medskip

The notes and pauses are viewed as compounds formed by a pitch (or lack of it, in the case of a pause) and a duration. For instance, in Figure \ref{fig:mission_input}, the first and the fourth notes are identical: `B\musFlat{}$_4$ quaver'. But the ninth and the eleventh are not: one is a `C$_4$ dotted minim' and the other is a `C$_4$ crotchet'. To avoid confusion with standard music theory, from now on, a note or pause will be referred to as an \textit{event}.

\medskip

\begin{figure}[H]
\begin{center}\vspace{0.5cm}
\includegraphics[width=0.35\linewidth]{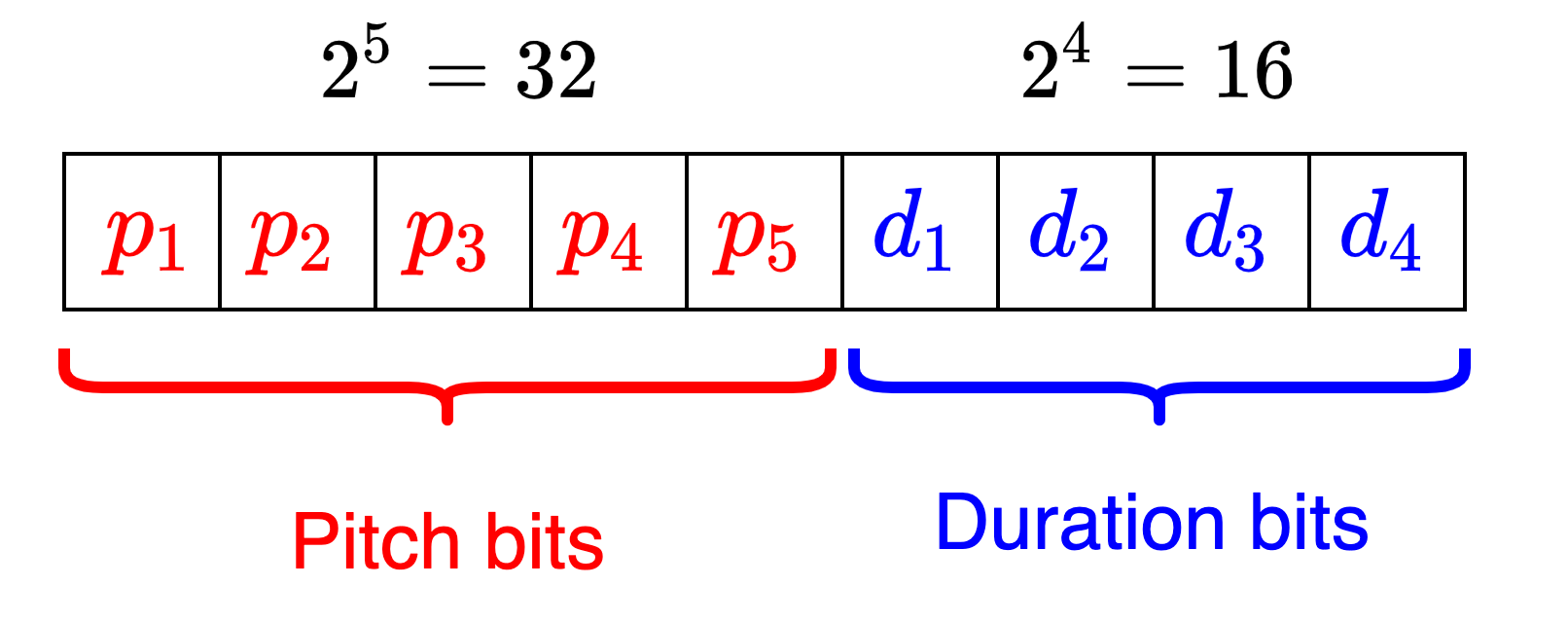}
\caption{Scheme for encoding the events of a musical sequence.}
\label{fig:iccmr_representation}
\end{center}
\end{figure}

An event is encoded using a string of nine binary digits (Figure \ref{fig:iccmr_representation}). The first five digits encode its pitch and the subsequent four encode its duration, that is, a rhythmic figure, such as minim, crotchet, quaver, and so on. Therefore, the system can process up to 32 distinct pitches (one of which is silence) and up to 16 distinct durations in a piece of music. 

\medskip

For the input example in Figure \ref{fig:mission_input}, the system identified six pitches and four rhythmic figures, and calculated the corresponding 5-bit and 4-bit codes shown in Tables \ref{table:pitch_bits}  and \ref{table:duration_bits}, respectively.

\begin{table}[H]
\centering
\begin{tabular}{|l|l|l|l|l|l|l|l|l|l|l|l|l|}
\hline
\multicolumn{1}{|c|}{\textbf{Note}} & \multicolumn{1}{c|}{\textbf{MIDI}}  & \multicolumn{1}{c|}{\textbf{5-bit code}} \\ \hline
\multicolumn{1}{|c|}{Silence} & \multicolumn{1}{c|}{0} & \multicolumn{1}{c|}{00000} \\ \hline
\multicolumn{1}{|c|}{B\musFlat{}$_3$} & \multicolumn{1}{c|}{58} & \multicolumn{1}{c|}{00001}  \\ \hline
\multicolumn{1}{|c|}{C$_4$} & \multicolumn{1}{c|}{60} & \multicolumn{1}{c|}{00010}  \\ \hline
\multicolumn{1}{|c|}{C\musSharp$_4$} & \multicolumn{1}{c|}{61} & \multicolumn{1}{c|}{00011}  \\ \hline
\multicolumn{1}{|c|}{D$_4$} & \multicolumn{1}{c|}{62} & \multicolumn{1}{c|}{00100} \\ \hline
\multicolumn{1}{|c|}{G$_4$} & \multicolumn{1}{c|}{67} & \multicolumn{1}{c|}{00101}  \\ \hline
\multicolumn{1}{|c|}{B\musFlat{}$_4$} & \multicolumn{1}{c|}{70} & \multicolumn{1}{c|}{00110}  \\ \hline
\end{tabular}
\caption{\footnotesize{Bit codes for pitches. MIDI note number equal to 0 corresponds to silence; i.e., no pitch.}}
\label{table:pitch_bits}
\end{table}

\begin{table}[H]
\centering
\begin{tabular}{|l|l|l|l|l|l|l|l|l|l|l|l|l|}
\hline
\multicolumn{1}{|c|}{\textbf{Duration}} & \multicolumn{1}{c|}{\textbf{Ticks}}  & \multicolumn{1}{c|}{\textbf{4-bit code}} \\ \hline
\multicolumn{1}{|c|}{quaver} & \multicolumn{1}{c|}{480} & \multicolumn{1}{c|}{0000} \\ \hline
\multicolumn{1}{|c|}{crotchet} & \multicolumn{1}{c|}{960} & \multicolumn{1}{c|}{0001}  \\ \hline
\multicolumn{1}{|c|}{minim + crotchet} & \multicolumn{1}{c|}{2400} & \multicolumn{1}{c|}{0010}  \\ \hline
\multicolumn{1}{|c|}{minim + quaver }& \multicolumn{1}{c|}{2880} & \multicolumn{1}{c|}{0011}  \\ \hline
\end{tabular}
\caption{\footnotesize{Bit codes for durations. Ticks are expressed in terms of the PPQN (Pulses per Quarter Note) time resolution of the MIDI input; e.g., a crotchet is set to 960 ticks.}}
\label{table:duration_bits}
\end{table}

\medskip

In total there are 12 events in our input tune, represented as shown in Eq. \ref{eq:raw_input}. Let us refer to this as a training set $\mathcal{T}$. If there were more than one input file, then they would be organised as sub-sets of the overall training set $\mathcal{T}$.

\begin{equation}
\begin{matrix}
\mathcal{T} = \{\: 001100000, 001010000, 001000011, \\
001100000, 001010000, 000110011, \\
001100000, 001010000, 000100011, \\
000010000, 000100001, 000000010\: \}
\end{matrix}
\label{eq:raw_input}
\end{equation}

\subsection{Musical lexicon}
\begin{figure}[H]
\begin{center}\vspace{0.5cm}
\includegraphics[width=0.55\linewidth]{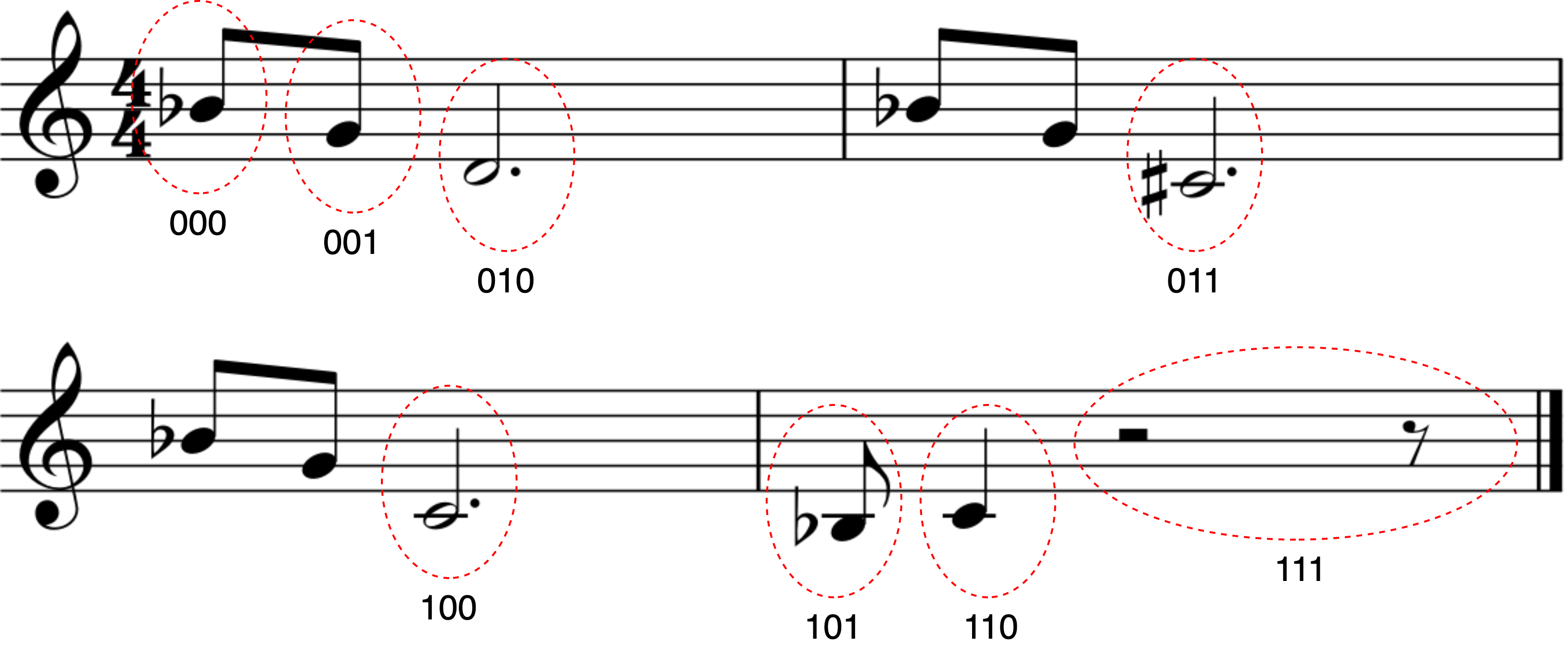}
\caption{The input tune has eight distinct events, highlighted in red.}
\label{fig:annotated_input}
\end{center}
\end{figure}

The next step builds a lexicon of unique events in $\mathcal{T}$. In practice, the system removes all repetitions of identical events. For this example, the lexicon $\mathcal{L}$ contains eight distinctive events, as shown in Eq. \ref{eq:distinctive_events}. Figure \ref{fig:annotated_input} highlights them on the score. (For now, ignore the binary codes written below them. These will be clarified below.)

\begin{equation}
\begin{matrix}
\mathcal{L} = \{\: 001100000, 001010000, 001000011, 000110011, \\
000100011, 000010000, 000100001, 000000010\: \}
\end{matrix}
\label{eq:distinctive_events}
\end{equation}

\subsection{Code compression}
\label{sec:code_compression}
At this stage, the system attempts to optimise, or compress, the lexicon, by relabelling its elements with, possibly, shorter binary codes. First, it estimates the number of bits needed to generate the new binary codes and then proceeds with the relabelling accordingly. 

\medskip
Code compression is an important step because the number of bits needed for the relabelling defines the number of \textit{qubits} that will be needed to build quantum circuits in the generative phase. In our example, three bits are sufficient to represent the eight events of the lexicon. Thus, a reduction from nine to three bits, as follows (also shown in  Figure \ref{fig:annotated_input}):

\begin{itemize}
\item 001100000 $\Longrightarrow$ 000
\item 001010000 $\Longrightarrow$ 001
\item 001000011 $\Longrightarrow$ 010
\item 000110011 $\Longrightarrow$ 011
\item 000100011 $\Longrightarrow$ 100
\item 000010000 $\Longrightarrow$ 101
\item 000100001 $\Longrightarrow$ 110
\item 000000010 $\Longrightarrow$ 111
\end{itemize}

\subsection{Identify sequences and build an occurrence matrix}

Next, the system builds an occurrence matrix with the number of times an event $\mathnormal{E_{t+1}}$ (listed on the top row) follows another one $\mathnormal{E_t}$ (listed on the first column on the left side). Figure \ref{fig:occurence_matrix} shows the occurrence matrix for the tune in Figure \ref{fig:mission_input}. For instance, the event `G$_4$ quaver' (001) followed the event  `B\musFlat{}$_4$ quaver' (000) three times. QuSing also considers $\mathnormal{E_t}$ as a sequence of $n$ events; e.g., rule \{000 001\} $\Rightarrow$ 010(33\%) $\lor$ 011(33\%) $\lor$ 100(34\%).

\begin{figure}[H]
\centering
\begin{tabular}{|l|l|l|l|l|l|l|l|l|l|l|l|l|}
\hline
\multicolumn{1}{|c|}{}       & \multicolumn{1}{c|}{\textbf{000}} & \multicolumn{1}{c|}{\textbf{001}} & \multicolumn{1}{c|}{\textbf{010}} & \multicolumn{1}{c|}{\textbf{011}} & \multicolumn{1}{c|}{\textbf{100}} & \multicolumn{1}{c|}{\textbf{101}} & \multicolumn{1}{c|}{\textbf{110}} & \multicolumn{1}{c|}{\textbf{111}}\\ \hline
\multicolumn{1}{|c|}{\textbf{000}} & \multicolumn{1}{c|}{} & \multicolumn{1}{c|}{3} & \multicolumn{1}{c|}{} & \multicolumn{1}{c|}{} & \multicolumn{1}{c|}{} & \multicolumn{1}{c|}{} & \multicolumn{1}{c|}{} & \multicolumn{1}{c|}{} \\ \hline
\multicolumn{1}{|c|}{\textbf{001}} & \multicolumn{1}{c|}{} & \multicolumn{1}{c|}{} & \multicolumn{1}{c|}{1} & \multicolumn{1}{c|}{1} & \multicolumn{1}{c|}{1} & \multicolumn{1}{c|}{} & \multicolumn{1}{c|}{} & \multicolumn{1}{c|}{} \\ \hline
\multicolumn{1}{|c|}{\textbf{010}} & \multicolumn{1}{c|}{1} & \multicolumn{1}{c|}{} & \multicolumn{1}{c|}{} & \multicolumn{1}{c|}{} & \multicolumn{1}{c|}{} & \multicolumn{1}{c|}{} & \multicolumn{1}{c|}{} & \multicolumn{1}{c|}{} \\ \hline
\multicolumn{1}{|c|}{\textbf{011}} & \multicolumn{1}{c|}{1} & \multicolumn{1}{c|}{} & \multicolumn{1}{c|}{} & \multicolumn{1}{c|}{} & \multicolumn{1}{c|}{} & \multicolumn{1}{c|}{} & \multicolumn{1}{c|}{} & \multicolumn{1}{c|}{} \\ \hline
\multicolumn{1}{|c|}{\textbf{100}} & \multicolumn{1}{c|}{}  & \multicolumn{1}{c|}{} & \multicolumn{1}{c|}{} & \multicolumn{1}{c|}{} & \multicolumn{1}{c|}{}  & \multicolumn{1}{c|}{1} & \multicolumn{1}{c|}{}  & \multicolumn{1}{c|}{} \\ \hline
\multicolumn{1}{|c|}{\textbf{101}} & \multicolumn{1}{c|}{}  & \multicolumn{1}{c|}{} & \multicolumn{1}{c|}{}  & \multicolumn{1}{c|}{}  & \multicolumn{1}{c|}{}  & \multicolumn{1}{c|}{} & \multicolumn{1}{c|}{1}& \multicolumn{1}{c|}{} \\ \hline
\multicolumn{1}{|c|}{\textbf{110}} & \multicolumn{1}{c|}{} & \multicolumn{1}{c|}{} & \multicolumn{1}{c|}{} & \multicolumn{1}{c|}{} & \multicolumn{1}{c|}{} & \multicolumn{1}{c|}{} & \multicolumn{1}{c|}{} & \multicolumn{1}{c|}{1} \\ \hline
\multicolumn{1}{|c|}{\textbf{111}} & \multicolumn{1}{c|}{} & \multicolumn{1}{c|}{} & \multicolumn{1}{c|}{} & \multicolumn{1}{c|}{} & \multicolumn{1}{c|}{} & \multicolumn{1}{c|}{} & \multicolumn{1}{c|}{} & \multicolumn{1}{c|}{} \\ \hline
\end{tabular}
\caption{\footnotesize{Occurence matrix for the tune in Figure \ref{fig:mission_input}.}}
\label{fig:occurence_matrix}
\end{figure}

\medskip

Recall from section \ref{sec:quantum_gen} above that the quantum circuit to compute the next note needs the occurrence values specified in terms of amplitudes. Therefore, the numbers of occurrences in the matrix need to be converted into amplitudes. The converted matrix is shown in Figure \ref{fig:transition_amplitudes}. Note that the last row was removed because event 111 has no successor. Should this event be produced during a generative process (rule $110 \Longrightarrow 111 (100\%$), then QuSing will
add 111 to the new composition and generate the next event (pseudo-)randomly.

\begin{figure}[H]
\centering
\begin{tabular}{|l|l|l|l|l|l|l|l|l|l|l|l|l|}
\hline
\multicolumn{1}{|c|}{}       & \multicolumn{1}{c|}{\textbf{000}} & \multicolumn{1}{c|}{\textbf{001}} & \multicolumn{1}{c|}{\textbf{010}} & \multicolumn{1}{c|}{\textbf{011}} & \multicolumn{1}{c|}{\textbf{100}} & \multicolumn{1}{c|}{\textbf{101}} & \multicolumn{1}{c|}{\textbf{110}} & \multicolumn{1}{c|}{\textbf{111}}\\ \hline
\multicolumn{1}{|c|}{\textbf{000}} & \multicolumn{1}{c|}{0.0} & \multicolumn{1}{c|}{\color{red}{1.0}} & \multicolumn{1}{c|}{0.0} & \multicolumn{1}{c|}{0.0} & \multicolumn{1}{c|}{0.0} & \multicolumn{1}{c|}{0.0} & \multicolumn{1}{c|}{0.0} & \multicolumn{1}{c|}{0.0} \\ \hline
\multicolumn{1}{|c|}{\textbf{001}} & \multicolumn{1}{c|}{0.0} & \multicolumn{1}{c|}{0.0} & \multicolumn{1}{c|}{\color{red}{0.58}} & \multicolumn{1}{c|}{\color{red}{0.58}} & \multicolumn{1}{c|}{\color{red}{0.58}} & \multicolumn{1}{c|}{0.0} & \multicolumn{1}{c|}{0.0} & \multicolumn{1}{c|}{0.0} \\ \hline
\multicolumn{1}{|c|}{\textbf{010}} & \multicolumn{1}{c|}{\color{red}{1.0}} & \multicolumn{1}{c|}{0.0} & \multicolumn{1}{c|}{0.0} & \multicolumn{1}{c|}{0.0} & \multicolumn{1}{c|}{0.0} & \multicolumn{1}{c|}{0.0} & \multicolumn{1}{c|}{0.0} & \multicolumn{1}{c|}{0.0} \\ \hline
\multicolumn{1}{|c|}{\textbf{011}} & \multicolumn{1}{c|}{\color{red}{1.0}} & \multicolumn{1}{c|}{0.0} & \multicolumn{1}{c|}{0.0} & \multicolumn{1}{c|}{0.0} & \multicolumn{1}{c|}{0.0} & \multicolumn{1}{c|}{0.0} & \multicolumn{1}{c|}{0.0} & \multicolumn{1}{c|}{0.0} \\ \hline
\multicolumn{1}{|c|}{\textbf{100}} & \multicolumn{1}{c|}{0.0}  & \multicolumn{1}{c|}{0.0} & \multicolumn{1}{c|}{0.0} & \multicolumn{1}{c|}{0.0} & \multicolumn{1}{c|}{0.0}  & \multicolumn{1}{c|}{\color{red}{1.0}} & \multicolumn{1}{c|}{0.0}  & \multicolumn{1}{c|}{0.0} \\ \hline
\multicolumn{1}{|c|}{\textbf{101}} & \multicolumn{1}{c|}{0.0}  & \multicolumn{1}{c|}{0.0} & \multicolumn{1}{c|}{0.0}  & \multicolumn{1}{c|}{0.0}  & \multicolumn{1}{c|}{0.0}  & \multicolumn{1}{c|}{0.0} & \multicolumn{1}{c|}{\color{red}{1.0}}& \multicolumn{1}{c|}{0.0} \\ \hline
\multicolumn{1}{|c|}{\textbf{110}} & \multicolumn{1}{c|}{0.0} & \multicolumn{1}{c|}{0.0} & \multicolumn{1}{c|}{0.0} & \multicolumn{1}{c|}{0.0} & \multicolumn{1}{c|}{0.0} & \multicolumn{1}{c|}{0.0} & \multicolumn{1}{c|}{0.0} & \multicolumn{1}{c|}{\color{red}{1.0}} \\ \hline
\end{tabular}
\caption{\footnotesize{Occurrence matrix with amplitudes, highlighted in red. The last row of matrix was removed.}}
\label{fig:transition_amplitudes}
\end{figure}

\subsection{Pick an initial event and initiate the generative process}

\begin{figure}[H]
\begin{center}\vspace{0.5cm}
\includegraphics[width=0.2\linewidth]{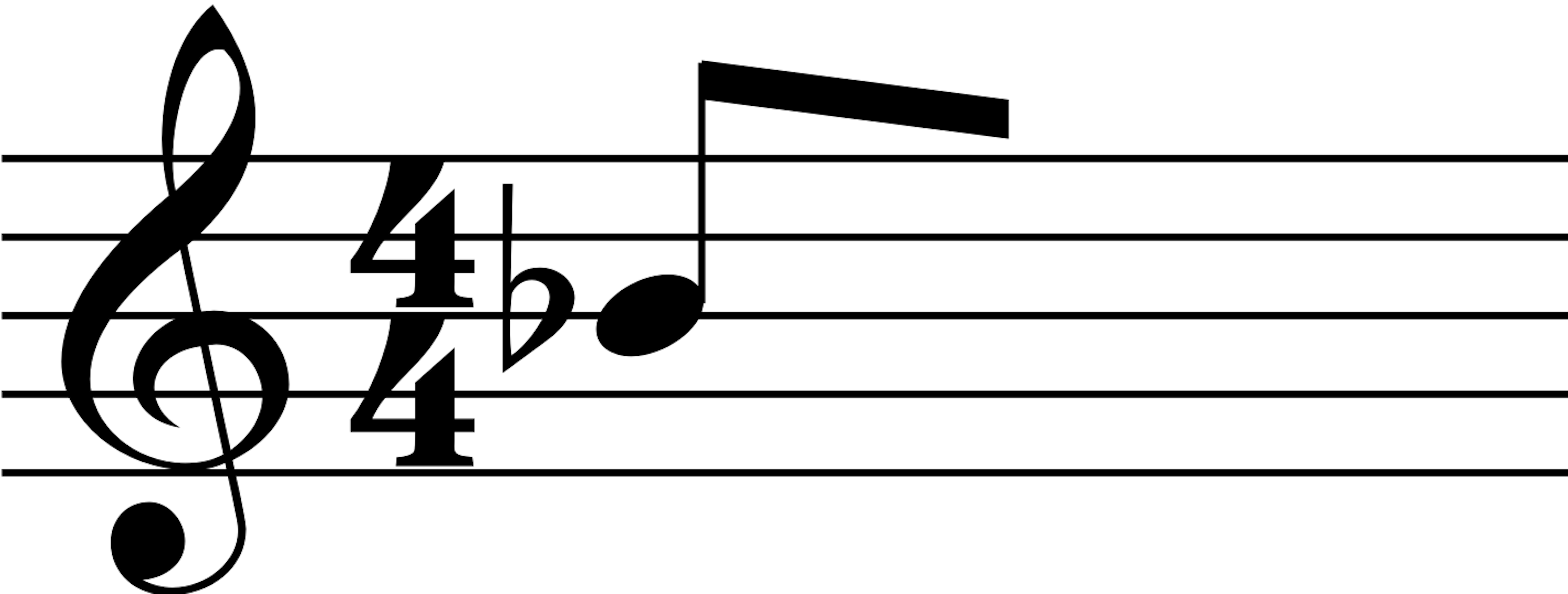}
\caption{The initial note.}
\label{fig:first_note}
\end{center}
\end{figure}

Now, the system picks an event $\mathnormal{E_t}$ to begin the generative process. By default, this is normally picked (pseudo-)randomly. But there is the option to start with the same beginning as the input piece (or from one of the input pieces, in case of using a corpus). For this example, we chose to begin with the input's first event: `B\musFlat{}$_4$ quaver' (000) (Figure \ref{fig:first_note}). The system is now ready to initiate the generative process.

\subsection{Build quantum circuit and measure}
\label{sec:build_and_measure}

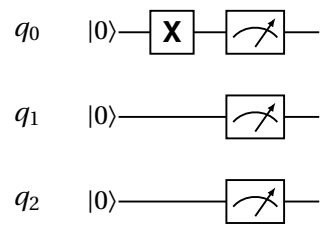
\begin{figure}[H]
\begin{center}\vspace{0.3cm}
    \begin{tikzpicture}
        \node[scale=0.9] 
        {
            \begin{quantikz}
                \lstick{$q_0$} & \ket{0} & \gate{\textbf{X}} &  \meter{}  & \qw  \\
                \lstick{$q_1$} & \ket{0} & \qw & \meter{}  & \qw  \\
                \lstick{$q_2$} & \ket{0} & \qw & \meter{} & \qw 
                \lstick{}
                \lstick{\\}
            \end{quantikz}
        };  
    \end{tikzpicture}
\end{center}
\caption{Circuit for the transition \textbf{000} $\Longrightarrow$ [0.0, {\color{red}1.0}, 0.0, 0.0, 0.0, 0.0, 0.0, 0.0].}
\label{fig:simple_circuit}
\end{figure}

\begin{figure}[H]
\begin{center}\vspace{0.5cm}
\includegraphics[width=0.2\linewidth]{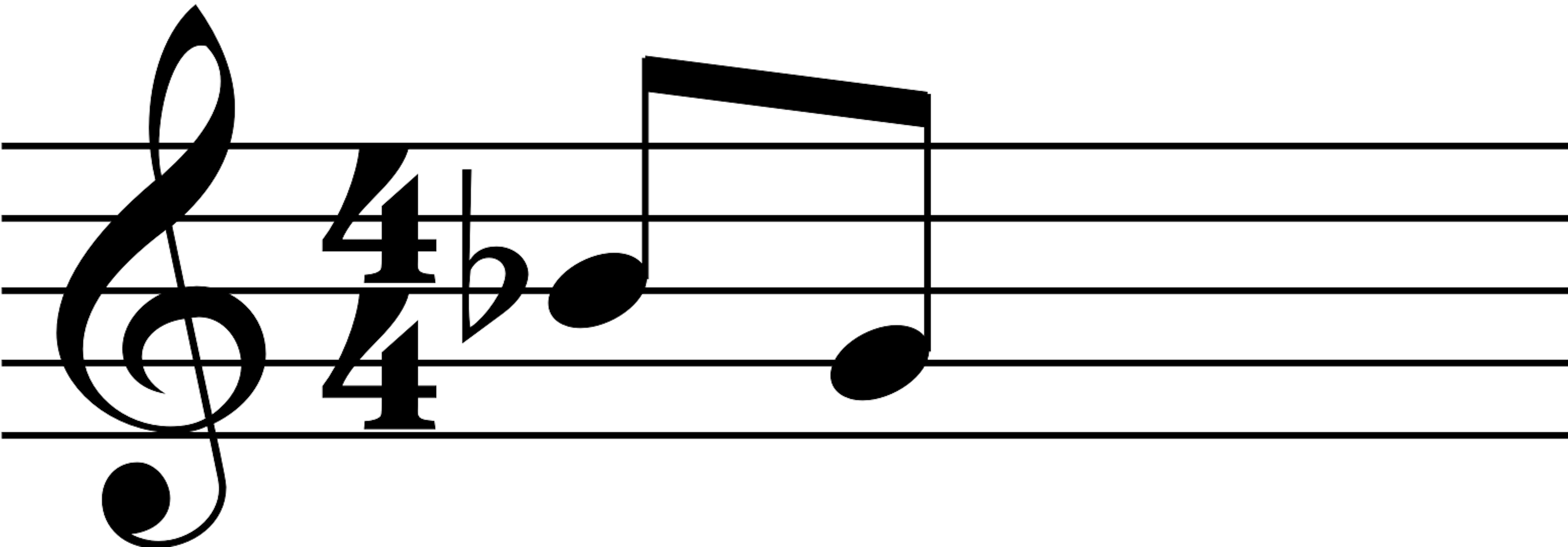}
\caption{The resuting second note is appended to the sequence.}
\label{fig:second_note}
\end{center}
\end{figure}

According to our occurrence matrix (Figure \ref{fig:transition_amplitudes}), only event 001 can follow event 000. The respective amplitude for 001 is equal to 1.0, which means that there is a 100\% probability that 000 will be followed by 001 (Figure \ref{fig:second_note}). The quantum circuit\footnote{In our implementation, we used Qiskit's \texttt{initialize()} function to build the circuit.} to compute this transition is rather simple (Figure \ref{fig:simple_circuit}): it should always measure $q_2 = 0$, $q_1 = 0$, $q_0 = 1$, that is 001. In practice, as we are dealing with a small occurrence matrix in this example, here we simply skip the quantum processing altogether, and retrieve the only possible choice classically; see discussion in section \ref{sec:discussion}.

\medskip

At this point, the result $\mathnormal{E_{t+1}}$ = 001 is appended into the new tune, $\mathnormal{E}$ becomes equal to 001, and the system moves on to generate the next event.

\medskip

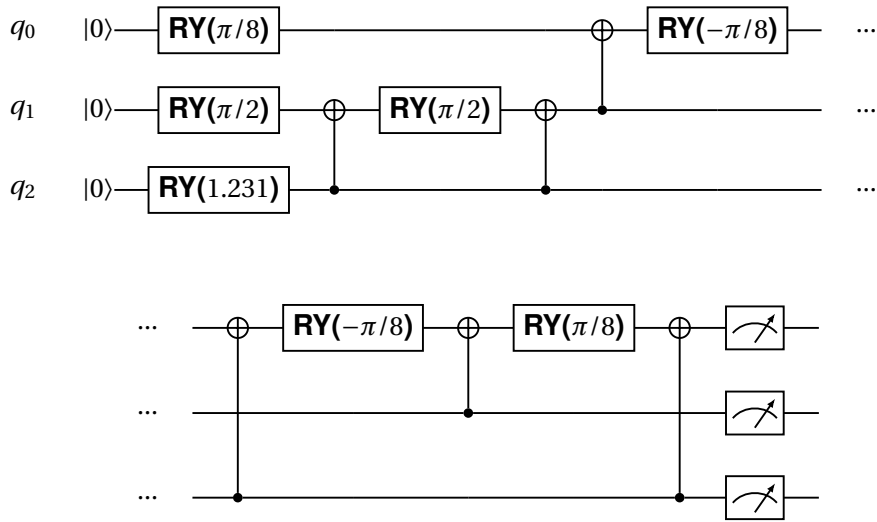
\begin{figure}[H]
\begin{center}
\vspace{0.3cm}
    \begin{tikzpicture}
        \node[scale=0.9] 
        {
            \begin{quantikz}
                \lstick{$q_0$} & \ket{0} & \gate{\textbf{RY($\pi/8$)}} & \qw & \qw & \qw  & \targ{} & \gate{\textbf{RY($-\pi/8$)}}  & \qw & ... & \\
                \lstick{$q_1$} & \ket{0} & \gate{\textbf{RY($\pi/2$)}} & \targ{} & \gate{\textbf{RY($\pi/2$)}} & \targ{} & \ctrl{-1} & \qw  & \qw & ... &\\
                \lstick{$q_2$} & \ket{0} & \gate{\textbf{RY($1.231$)}} & \ctrl{-1} & \qw & \ctrl{-1} & \qw & \qw  & \qw & ... & \\
                \lstick{} 
                \lstick{\\}
            \end{quantikz}
        };  
    \end{tikzpicture}
    
    \begin{tikzpicture}
        \node[scale=0.9] 
        {
            \begin{quantikz}
                \lstick{} & ... & & \targ{}  & \gate{\textbf{RY($-\pi/8$)}} & \targ{} & \gate{\textbf{RY($\pi/8$)}} & \targ{} & \meter{} & \qw  \\
                \lstick{} & ... & & \qw & \qw & \ctrl{-1} & \qw & \qw  & \meter{} & \qw \\
                \lstick{} & ... & & \ctrl{-2} & \qw & \qw & \qw & \ctrl{-2}  & \meter{} & \qw
            \end{quantikz}
        };
        
    \end{tikzpicture}
\end{center}
\caption{Circuit for the transition \textbf{001} $\Longrightarrow$ [0.0, 0.0, {\color{red}0.58}, {\color{red}0.58}, {\color{red}0.58}, 0.0, 0.0, 0.0].}
\label{fig:quantum-circuit}
\end{figure}

\begin{figure}[H]
\begin{center}\vspace{0.5cm}
\includegraphics[width=0.2\linewidth]{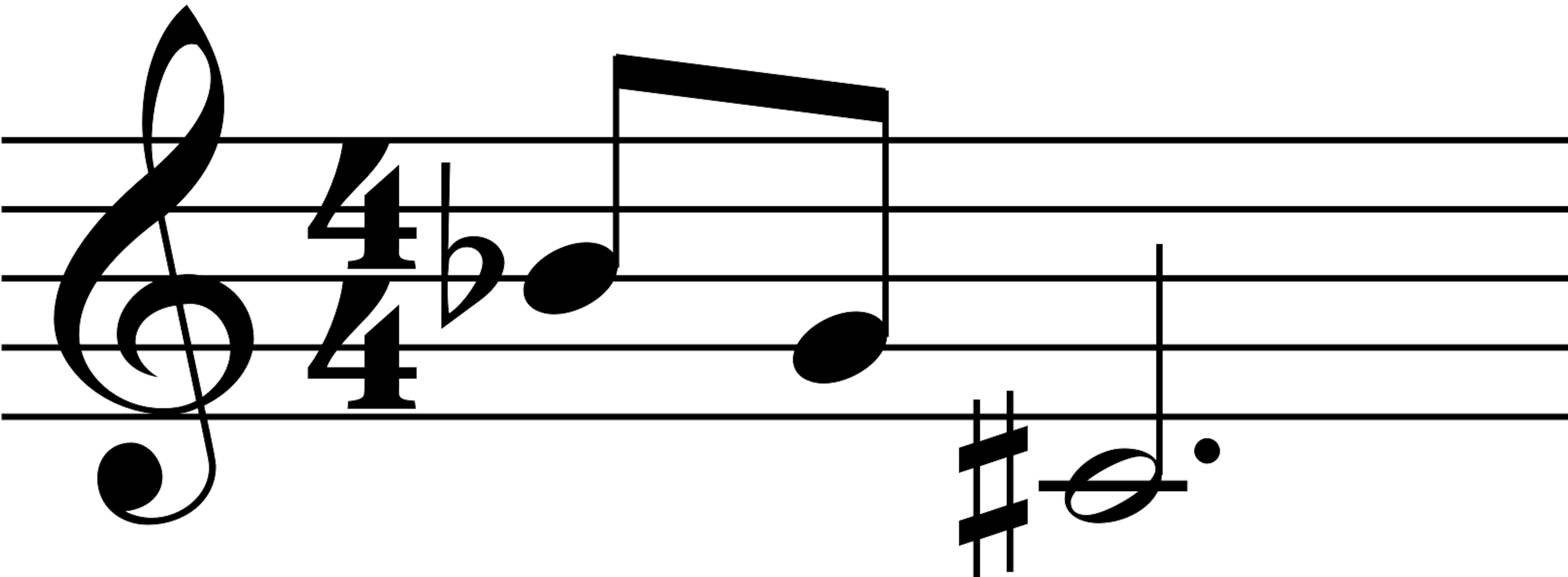}
\caption{The resuting third note is appended to the sequence.}
\label{fig:third_note}
\end{center}
\end{figure}

According to our occurrence matrix (Figure \ref{fig:transition_amplitudes}), three events have equal probability of 33\% (${0.58}^2 \times 100 = 33$) to follow 001: 010, 011 and 100. In this case, the system builds the circuit shown in Figure \ref{fig:quantum-circuit}. The respective measurement is then rendered into music and the process continues accordingly. In this example, the measurement returned 011, which corresponds to the event `C\musSharp$_4$ dotted minim' (Figure \ref{fig:third_note}).

\subsection{Decode musical event and synthesise singing}

\begin{figure}[H]
\begin{center}\vspace{0.8cm}
\includegraphics[width=0.5\linewidth]{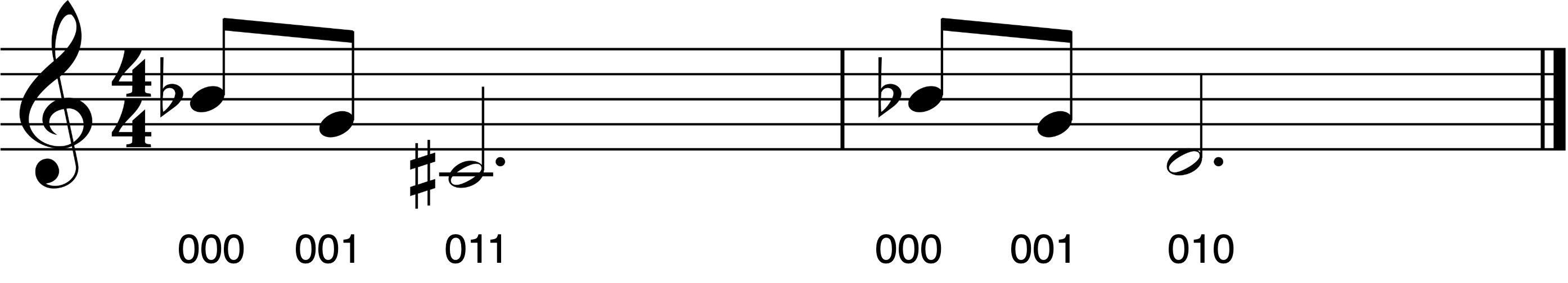}
\caption{Generated output after five generative cycles.}
\label{fig:example_output}
\end{center}
\end{figure}

\begin{figure}[H]
\begin{center}\vspace{0.8cm}
\includegraphics[width=0.6\linewidth]{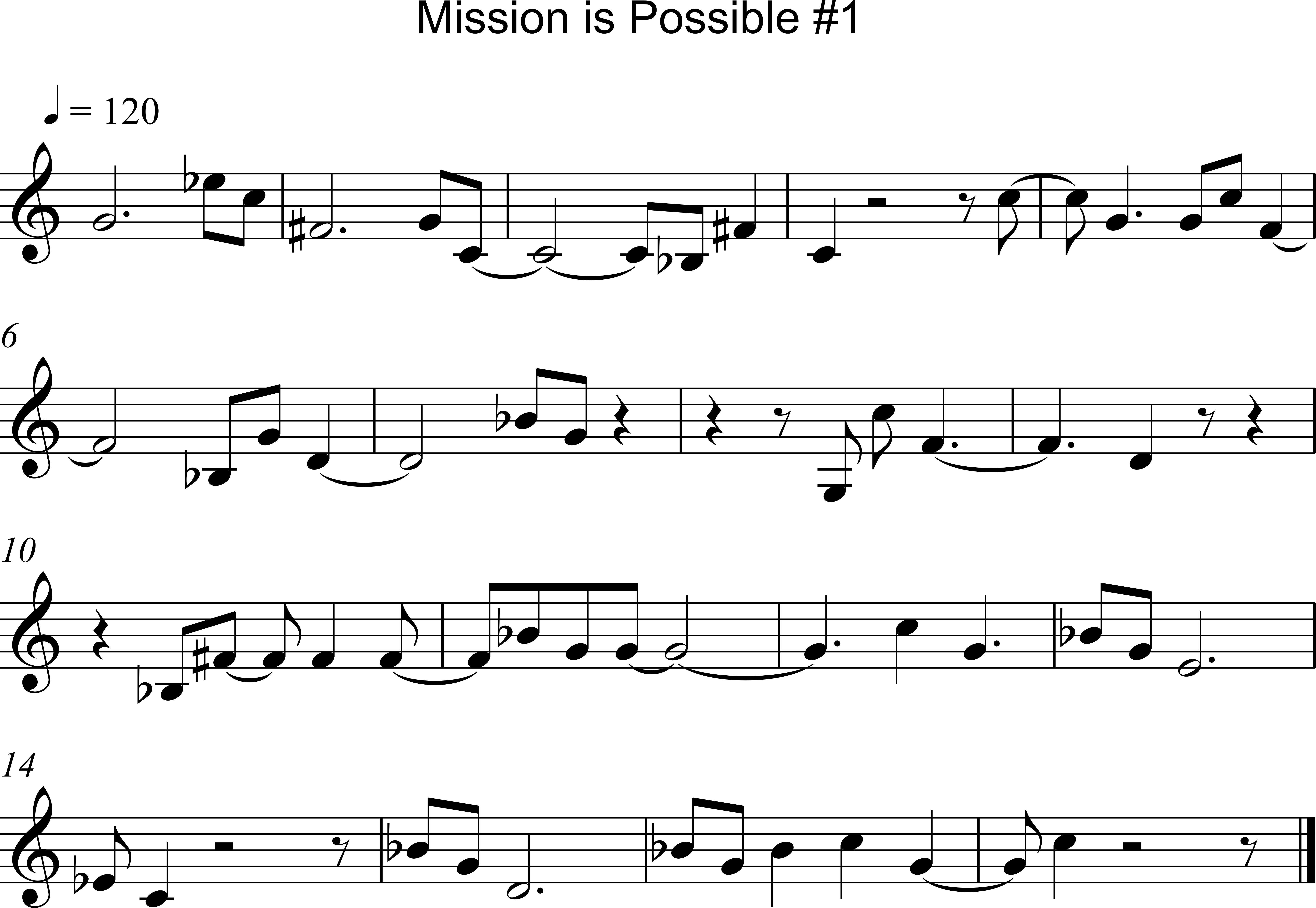}
\caption{QuSing-generated tune, with 50 rounds, 1 shot each.}
\label{fig:mission_qugen_1}
\end{center}
\end{figure}

Figure \ref{fig:example_output} shows the resulting tune after five generative iterations; see Figure \ref{fig:mission_qugen_1} and the Appendix for more examples. Once a measurement is obtained, the value is used to retrieve the respective `uncompressed` code (see section \ref{sec:code_compression}), which is then decoded according to Tables \ref{table:pitch_bits} and \ref{table:duration_bits}, respectively.

\medskip

The decoded event is either synthesised on the spot or appended into a MIDI file, which can be synthesised in batch mode after the process is terminated, or both. The synthesised tune is also stored either way. 

\begin{figure}[H]
\begin{center}\vspace{0.8cm}
\includegraphics[width=0.6\linewidth]{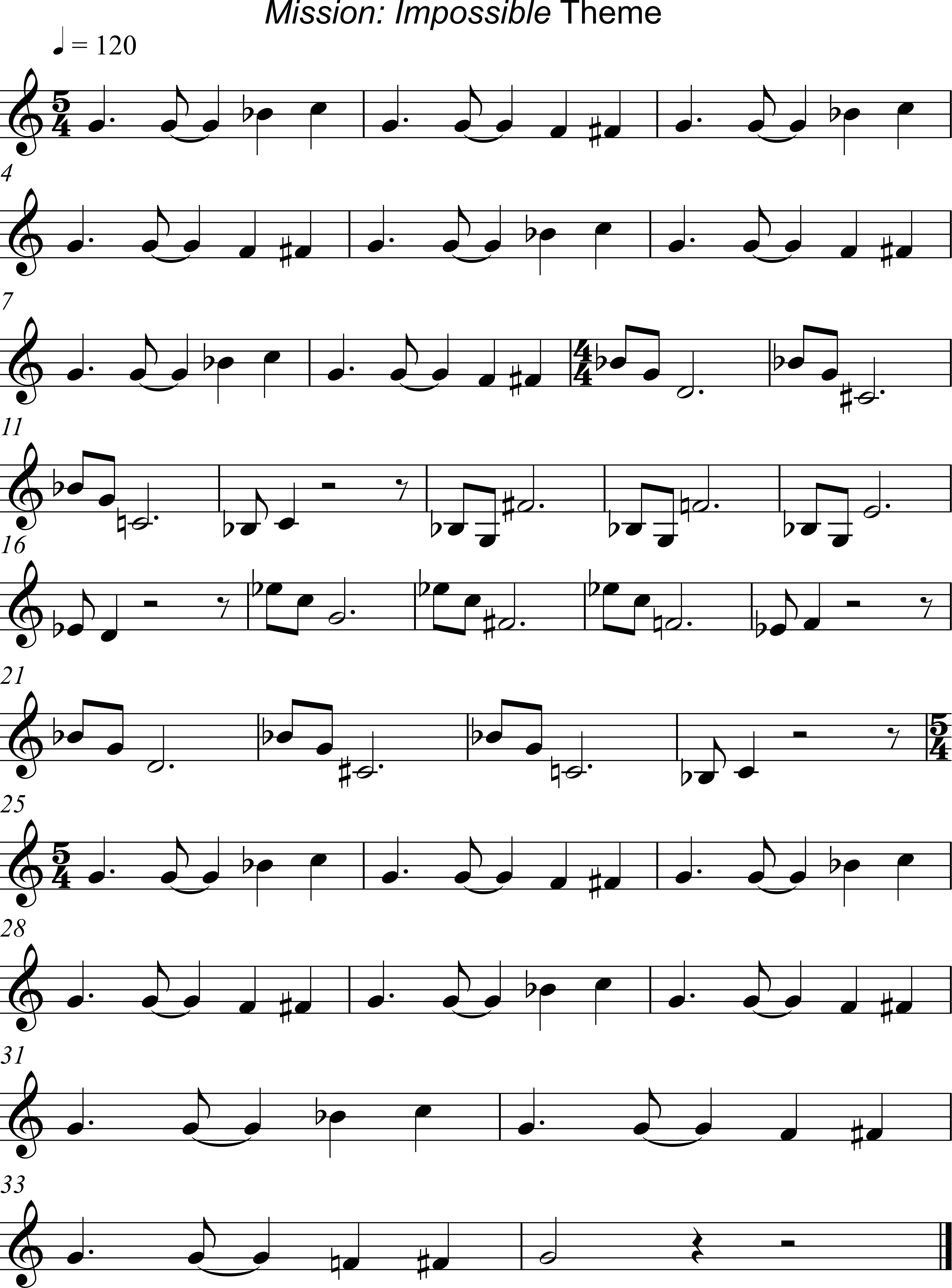}
\caption{\textit{Mission: Impossible} theme.}
\label{fig:mission_original}
\end{center}
\end{figure}

\medskip

Figure \ref{fig:mission_original} shows a longer version of the \textit{Mission: Impossible} theme input and a respective QuSing-generated version is presented in Figure \ref{fig:mission_qugen_1}. This was generated with IBM Quantum's \texttt{ibmq\_lima} backend, which is a 5-qubit superconductor processor. Here we asked QuSing to produce 50 events, with one shot per each event\footnote{Number of shots means how many times a circuit is run to get a probability distribution of results; see section \ref{sec:discussion} Discussion.}. 

%
%

\section{Discussion}
\label{sec:discussion}

We are interested in developing quantum computing systems for musical composition that supports the user's creative process by enabling experiments with different settings to produce varied outcomes \cite{Miranda2021b}. In this section, we examine QuSing's behaviour under different conditions and discuss how this may be harnessed for musical experimentation.

\subsection{Embracing noise}

Currently, quantum computers are crippled by noise, or decoherence \cite{Rieffel2011}: a quantum mechanics phenomenon that leads to processing errors.  Coherence time is the length of time qubits can hold quantum information. When qubits are disrupted by external interference, such as temperature changes or stray electromagnetic fields, information about the state of the qubits is destroyed. This can ruin the ability to exploit quantum mechanics for computation. Longer coherence times enable more quantum operations to be utilised before this occurs. Significant work is being conducted by the research community to mitigate decoherence problems. For instance, circuit optimisation techniques are aimed at reducing the number of gates to be executed on the qubits, thus reducing the time they are required to remain coherent. 

\medskip
For this project, we used Quantinuum's TKET optimisation tool \cite{QuantinuumTKET} and Qiskit's Pass Manager for circuit optimization \cite{QiskitPassManager} to make the circuits as small and efficient as possible. We then used the Python package \texttt{mapomatic} to optimize qubit assignment \cite{mapomatic}. This is useful because not all qubits are created equally and not all qubits are connected equally, so we choose the least-error-prone configuration. Finally, we used the Python package M3 (Matrix-free Measurement Mitigation, or \texttt{mthree}) to mitigate measurement-related errors \cite{M3}; it uses the quantum computer's calibration information to further reduce errors. In the near future, these low-level operational issues are likely to be transparent to the general user. And with the development of increasingly more fault-tolerant qubits and high connectivity\footnote{Gates involving more than one qubit need to be connected. But, not all qubits of a superconducting processor are connected, which limits the ability to entangle them.}, qubit assignment should not matter at a low level either.

\medskip
For a creative application, however, noise is not necessarily bad, provided it can be somehow controlled. Manageable degrees of noise can be useful, for example, to introduce surprises in the outcomes. In the case of QuSing, we can set it to tolerate `wrong' events under certain conditions. For example, it can tolerate an event that is not covered by the respective rule, provided that there exists a rule that would enable the generative process to continue from the wrong event. Otherwise, the result is discarded and the circuit is run again until it produces a permissible result. 

\subsection{Variability control}

Also useful, is the ability to control the degree of variation of an outcome. We can do this by changing the number of previous events in the rules. The higher the number $n$ of previous events, the higher the resemblance between the outputs and the training tune(s). 

\begin{figure}[H]
\begin{center}\vspace{0.8cm}
\includegraphics[width=0.5\linewidth]{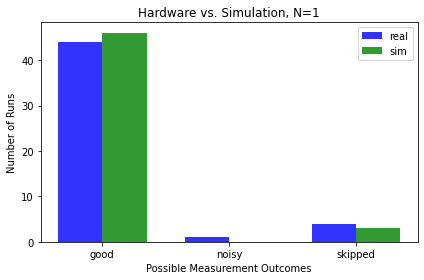}
\caption{Results from running QuTune for 50 rounds, with rules considering $n=1$ previous events.}
\label{fig:Bach_N=1}
\end{center}
\end{figure}

\begin{figure}[H]
\begin{center}\vspace{0.8cm}
\includegraphics[width=0.5\linewidth]{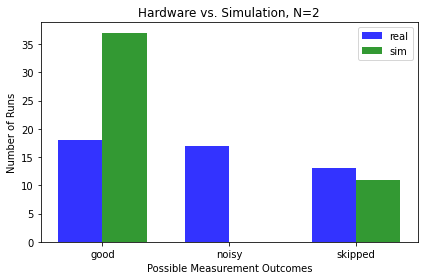}
\caption{Results from running QuTune for 50 rounds, with rules considering $n=2$ previous events.}
\label{fig:Bach_N=2}
\end{center}
\end{figure}

\begin{figure}[H]
\begin{center}\vspace{0.8cm}
\includegraphics[width=0.5\linewidth]{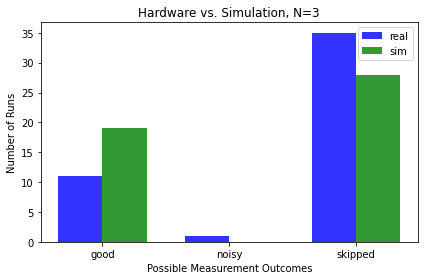}
\caption{Results from running QuTune for 50 rounds, with rules considering $n=3$ previous events.}
\label{fig:Bach_N=3}
\end{center}
\end{figure}

\medskip
Figures \ref{fig:Bach_N=1}, \ref{fig:Bach_N=2}, and \ref{fig:Bach_N=3} plot the results from running QuSing for 50 generative rounds, one shot each, with rules considering $n=1$, $n=2$, and $n=3$ previous events, respectively. We trained the system with an excerpt of J. S. Bach's \textit{Cello Suite Nr. 1} (Appendix \ref{sec:appendix2}, Figure \ref{fig:bach_orig}) to generate music on a simulator and on quantum hardware\footnote{For the simulations we used IBM Quantum's Aer simulators and for runs on real hardware we used the \texttt{ibmq\_lima} and \texttt{ibmq\_belem} backends.}. 

We monitored the variability of the results by counting the number of times the system skipped building a quantum circuit because there was only one possible choice, which was retrieved classically. There were fewer choices to be made with $n=3$ rules than with $n=1$ ones. Therefore the results with $n=1$ carry more variations than the results with $n=3$. For instance in Figures \ref{fig:Bach_N=1}, \ref{fig:Bach_N=2}, and \ref{fig:Bach_N=3}, the bars for `good' count the number of times the system built circuits, whereas the bars for `skipped' indicate when it has not done so. The `noisy` bar indicates the number of times QuSing produced a `wrong' event\footnote{In these examples the wrong events were not tolerated to produce the musical outputs shown in the Appendix.}; obviously, simulations are unlikely to produce wrong events. 


\medskip
Note, however, that in Figure \ref{fig:Bach_N=2} errors are prevalent. We were not able to establish the reason for this. One of the reasons for this discrepancy might be that we ran this on a different backend: for Figure \ref{fig:Bach_N=2} we used the \texttt{ibmq\_lima} backend, whereas for the other two we used the \texttt{ibmq\_belem} one. Surely, there is a better explanation, but this is not so relevant to scrutinise at this point. Sometimes we just get a bad start: we run it and it just keeps producing wrong events. At other times, if we get a good start, then it runs smoothly. Errors increase runtime because we have to wait in a queue each time we run a circuit. We might have to wait in multiple queues to get a good result. A queue could mean we wait for a minute, but it might mean we wait for many minutes. The respective musical results are available in Appendix, section \ref{sec:appendix2}. 

\subsection{Saving ammunition}

An error-mitigation method that is commonly applied in quantum computing is to run the same circuit many times and pick the result that occurs the most. These repetitions are referred to as \textit{shots}.

\begin{figure}[H]
\begin{center}\vspace{0.8cm}
\includegraphics[width=0.6\linewidth]{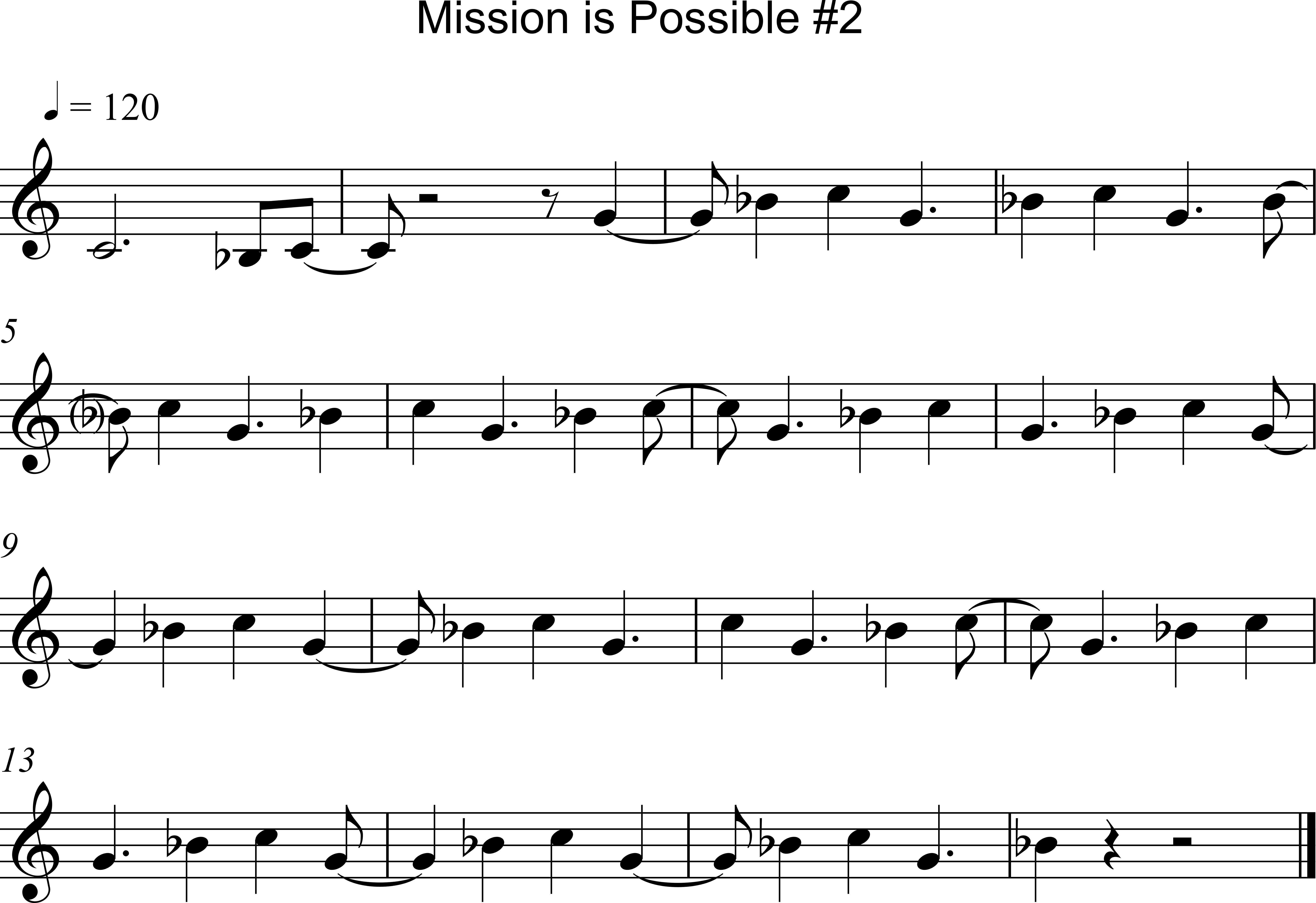}
\caption{QuSing-generated tune, with 50 rounds, 1,000 shots each.}
\label{fig:n=1000}
\end{center}
\end{figure}

\begin{figure}[H]
\begin{center}\vspace{0.8cm}
\includegraphics[width=0.5\linewidth]{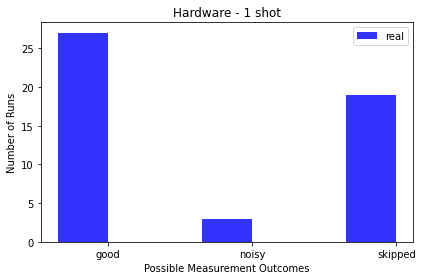}
\caption{Results from running the circuits for 1 shot.}
\label{fig:MI_1shot}
\end{center}
\end{figure}

\begin{figure}[H]
\begin{center}\vspace{0.8cm}
\includegraphics[width=0.5\linewidth]{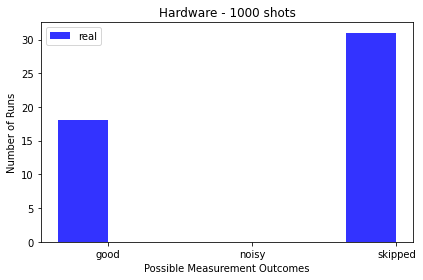}
\caption{Results from running the circuits for 1,000 shots.}
\label{fig:MI_1kshot}
\end{center}
\end{figure}

\medskip
In the case of QuSing, the number of shots can be used as a parameter to control the outcomes. For instance, consider one of the rules discussed in section \ref{sec:tran_rules}: $D_4 \Longrightarrow C_3(20\%) \lor E_3(80\%)$. Whereas running a circuit for this rule for many shots would minimise the chances of producing a wrong note (compare the `noise' bars in Figures  \ref{fig:MI_1shot} and \ref{fig:MI_1kshot}), it would significantly increase the chances of producing the note $E_3$. That is, running a circuit for many shots unbalances the respective rule towards the option that has the highest probability, excepting, of course, those rules with equal probabilities for all possible outcomes. As an example of the effect of the number of shots, compare the tunes generated with 1 shot per round shown in Figure \ref{fig:mission_qugen_1} and the one generated with 1,000 shots per round, with the same rules and backend, shown in Figure \ref{fig:n=1000}. In the latter, from the third musical bar onwards, the tune repeats the same motif until the end. In practice, the system got biased with only one option for most rules. This, combined with rules that already offered only one option anyway, caused the lack of variation.

\subsection{Classic, quantum, or hybrid?}

We mentioned earlier that, for rules that have only one possible outcome, the system skips the quantum processing and retrieves the only possible choice classically. For larger training inputs and with data of increased complexity, it is unlikely that there will be rules with only one outcome. On the contrary, we are likely to see rules with significantly more options with varied probabilities of occurrence. All the same, our choice of processing these cases classically is purely circumstantial, given the state of the art of technology we currently have access to. As discussed above, it would be far neater to process everything quantumly and benefit from quantum uncertainty and noise. But given the time it takes to communicate with a quantum machine over the cloud, and the fact that we currently have to do this for every new event, we might as well avoid doing this here.

\section{Final Remarks}
\label{sec:remarks}

\medskip
The field of quantum computing is incipient. The current stage of quantum computing technology is somewhat comparable with the early days of classical digital computers. They also were unportable, unreliable, high-maintenance, expensive, and required highly specialized skills to be handled. It is hard to imagine what computers will be like a few decades from now, in the same way that it must have been hard for our forefathers of the 1950s to imagine how computers ended up being like today. 

\medskip
It is fair to say that programming a quantum computer at the time of writing is comparable to having to program a digital computer using a low-level programming language: one has to write code that operates at the level of qubits. This is changing rapidly, as the industry is making impressive progress in developing high-level software development kits (SDK); e.g., Microsoft's QDK, IBM's Qiskit, Xanadu's Strawberry Fields, and Quantinuum's TKET, to cite but four.  Indeed, this paper discussed implementation and operational issues (e.g., noise, circuit optimisation, and sludgy Internet access to backends) that are likely to be transparent in the near future, but which, nevertheless, researchers developing the field should be aware of today. 

\medskip
At this stage, we are not in a position to advocate any quantum advantage for musical applications. What we advocate, however, is that the music technology community should be quantum-ready for when quantum computing hardware becomes more sophisticated, widely available, and possibly advantageous for creativity and business. In the process of learning and experimenting with this new technology, novel approaches, creative ideas, and innovative applications are bound to emerge. The method introduced here certainly is an example of an innovative approach to computing transition rules, which is truly \textit{quantum native}.

\medskip
By way of future work, we are in the process of increasing the amount of information that constitutes an event, which will include parameters for expressive singing. In the system presented in this paper, a musical event is constituted of two only kinds of information: pitch and duration. However, the vocal synthesisers that we are working with have several parameters to control vocal characteristics such as air inhalation and exhalation, lung pressure, mouth opening, vibrato, nasalisation, and attack time (for consonants), and more. This will require longer bit strings to represent the events, and consequently more qubits and circuits of increased complexity. 

\vspace{5mm} 

\textbf{Acknowledgements}

\medskip
This work was developed as part of the QuTune Project\footnote{https://iccmr-quantum.github.io/}, funded by the UK National Quantum Technologies Programme’s QCS Hub. For the purpose of open access, the authors have applied a Creative Commons Attribution (CC BY) licence to any Author Accepted Manuscript version arising.

\newpage

\section{Appendix}
\label{sec:appendix}
Audio recordings are available \cite{MusicExamples, QuSingQuTune}.

\subsection{\textit{Mission: Impossible} Experiments}
\label{sec:appendix1}

\begin{figure}[H]
\begin{center}\vspace{0.8cm}
\includegraphics[width=0.6\linewidth]{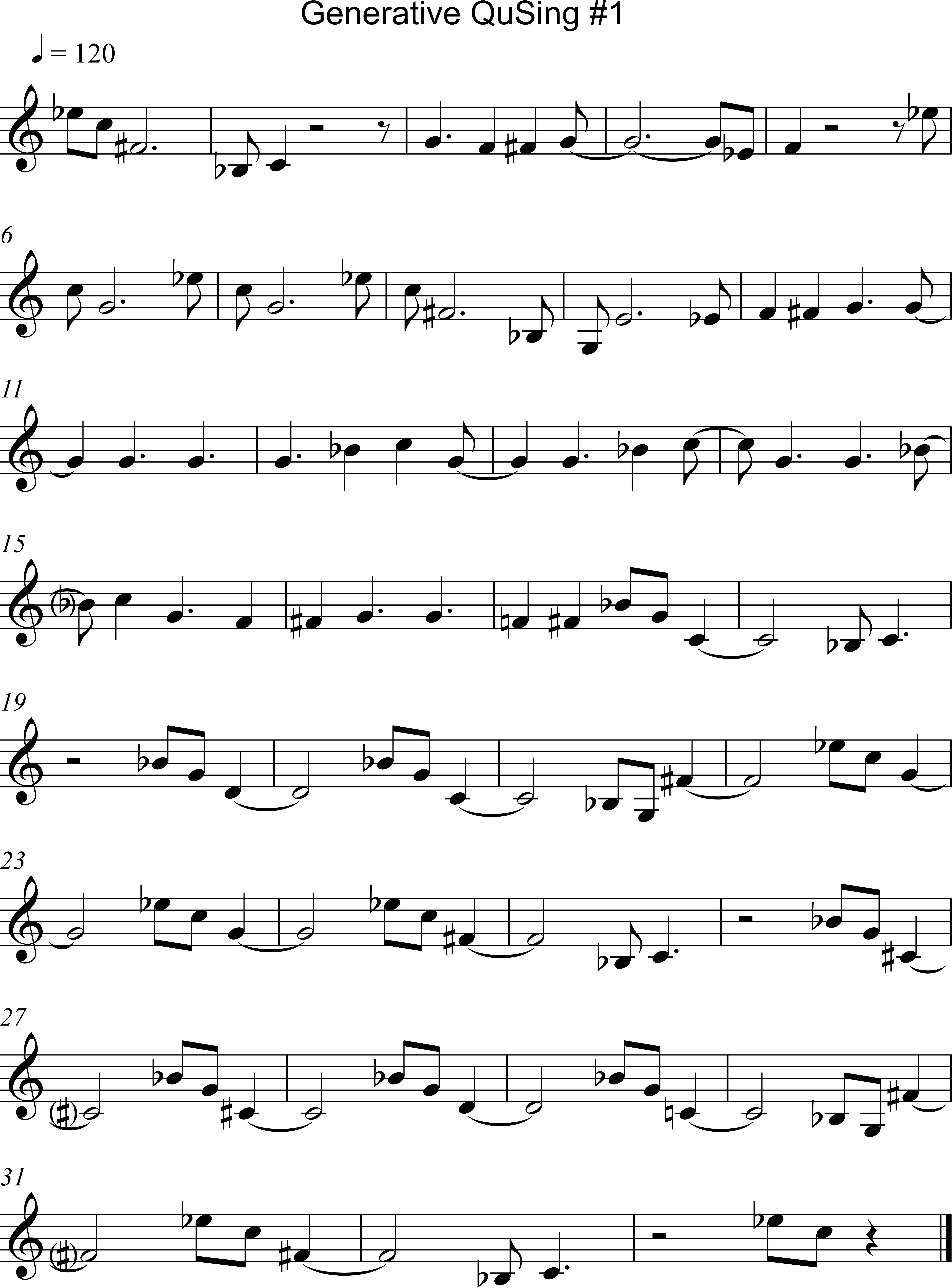}
\caption{Music generated with $n=1$.}
\label{fig:mi_simul_n=1}
\end{center}
\end{figure}

\begin{figure}[H]
\begin{center}\vspace{0.8cm}
\includegraphics[width=0.6\linewidth]{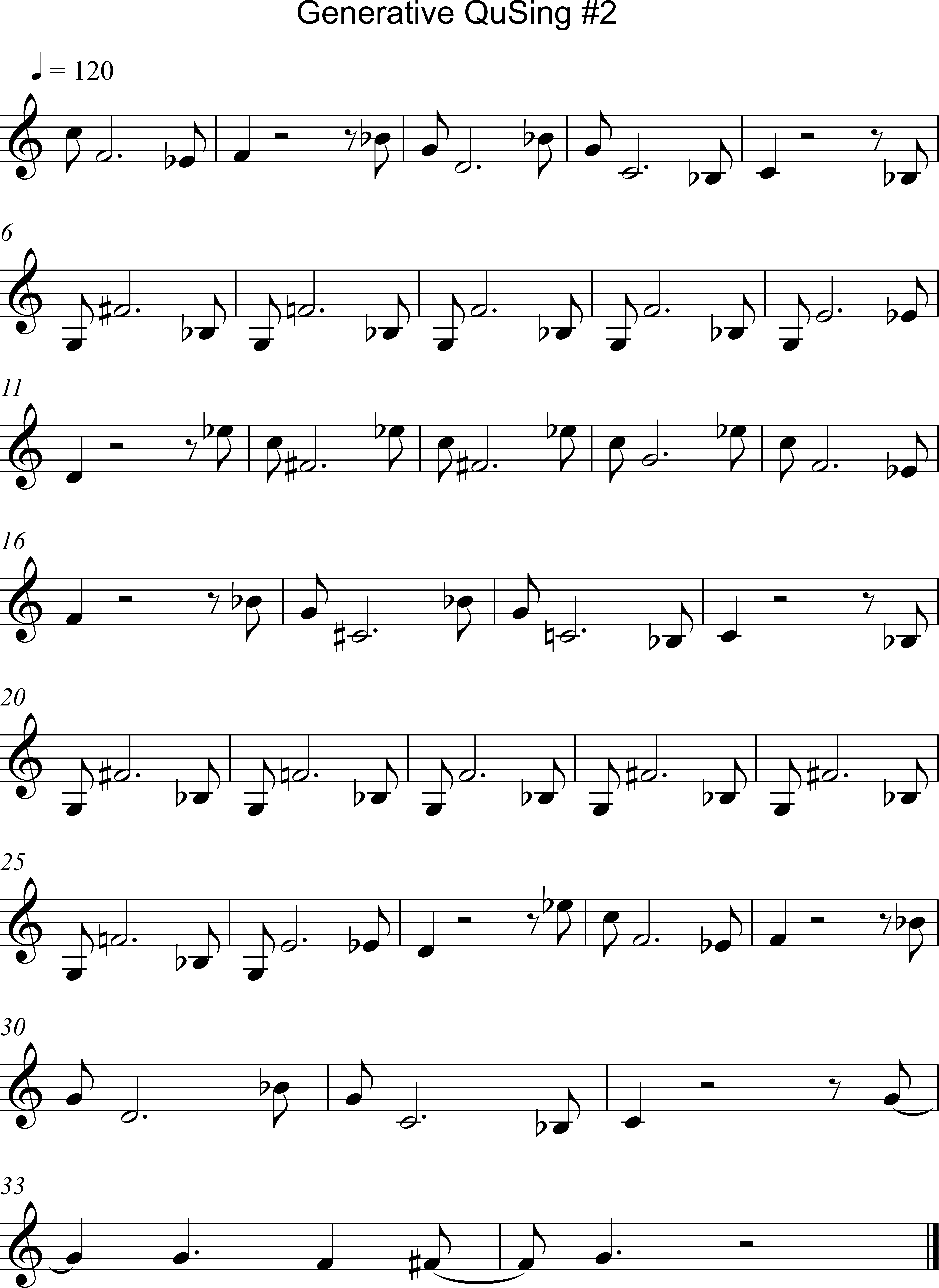}
\caption{Music generated with $n=2$.}
\label{fig:mi_simul_n=2}
\end{center}
\end{figure}

\begin{figure}[H]
\begin{center}\vspace{0.8cm}
\includegraphics[width=0.6\linewidth]{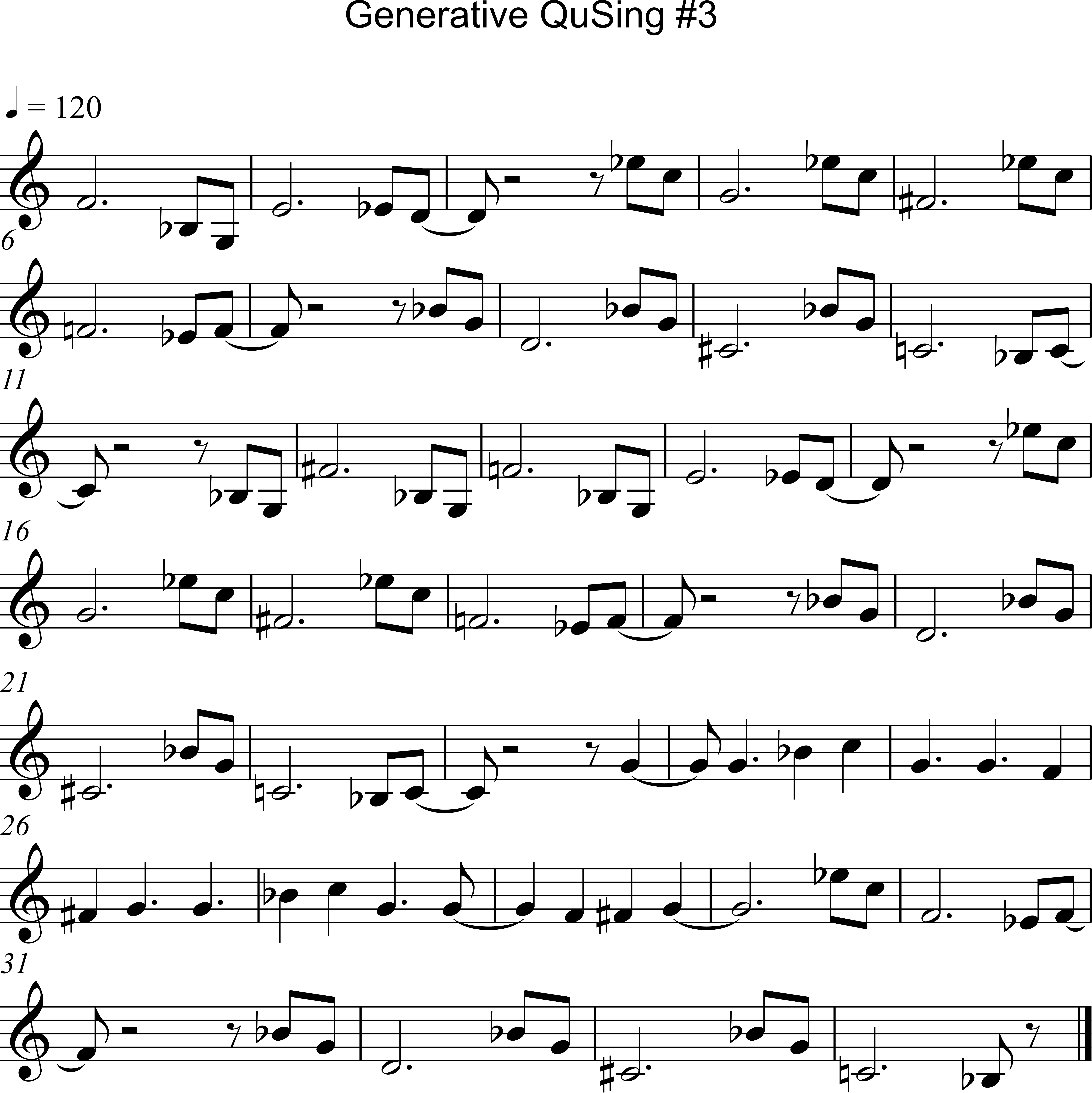}
\caption{Music generated with $n=3$.}
\label{fig:mi_simul_n=3}
\end{center}
\end{figure}

Given the \textit{Mission: Impossible} theme shown in Figure \ref{fig:mission_original} to train QuSing, Figures \ref{fig:mi_simul_n=1}, \ref{fig:mi_simul_n=2}, \ref{fig:mi_simul_n=3} shows three examples of outputs generated with $n = 1$, $n = 2$, and $n = 3$, where $n$ is the number of previous events in the rules. These were run on IBM Quantum's Aer simulator. The system was set to produce 100 events with one shot per generative cycle.

\newpage

\subsection{Q. C. Bach Experiments: Hardware vs. Simulations}
\label{sec:appendix2}

\begin{figure}[H]
\begin{center}\vspace{0.8cm}
\includegraphics[width=0.6\linewidth]{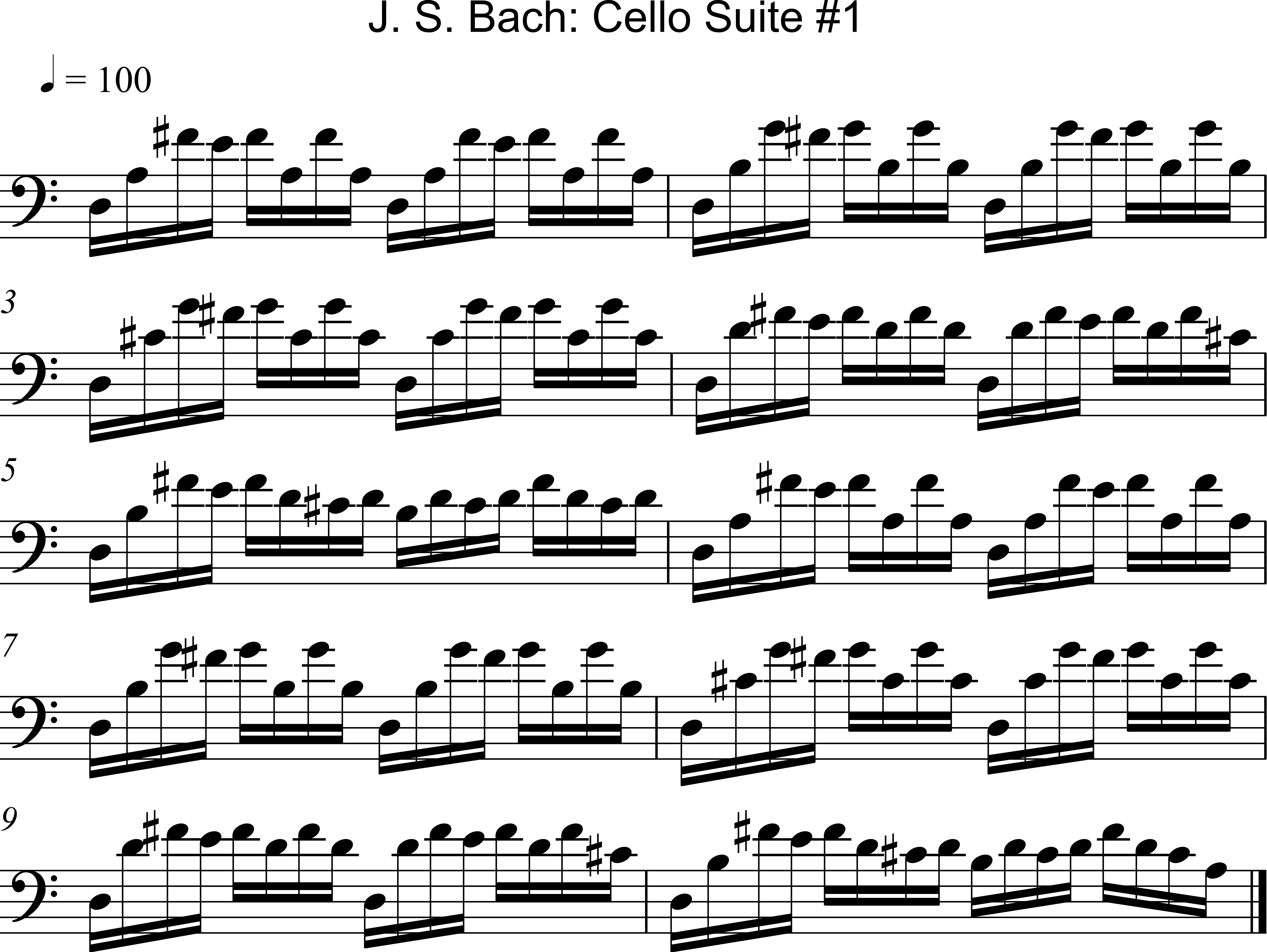}
\caption{Original excerpt of J. S. Bach's \textit{Cello Suite Nr. 1}.}
\label{fig:bach_orig}
\end{center}
\end{figure}

Below are results from experiments with hardware and simulation, with $n = 1$, $n = 2$, and $n = 3$, where $n$ is the number of previous events in the rules. The input music is an excerpt from J. S. Bach's \textit{Cello Suite Nr. 1}, which is shown in Figure \ref{fig:bach_orig}. The outputs are from running the system to generate 50 events. Note that the system increased the duration of the notes (semiquavers became crotchets) to make the tunes more fitting for singing. For the simulations, we used IBM Quantum's Aer simulator. For quantum hardware, we used \texttt{ibmq\_lima} and \texttt{ibmq\_belem} backends.

\begin{figure}[H]
\begin{center}\vspace{0.8cm}
\includegraphics[width=0.6\linewidth]{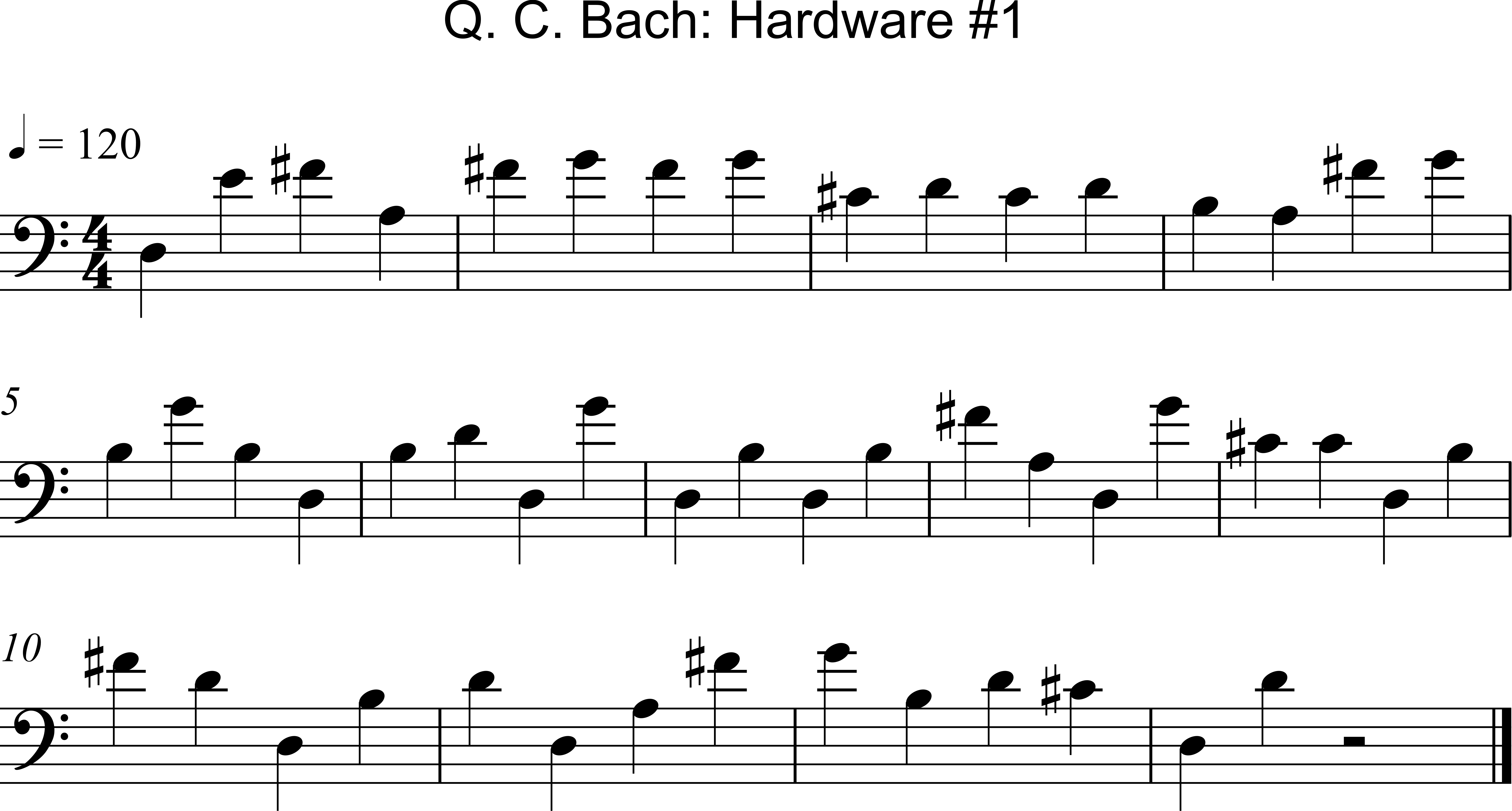}
\caption{Quantum generated Bach, hardware, $n = 1$.}
\label{fig:bach_hard1}
\end{center}
\end{figure}

\begin{figure}[H]
\begin{center}\vspace{0.8cm}
\includegraphics[width=0.6\linewidth]{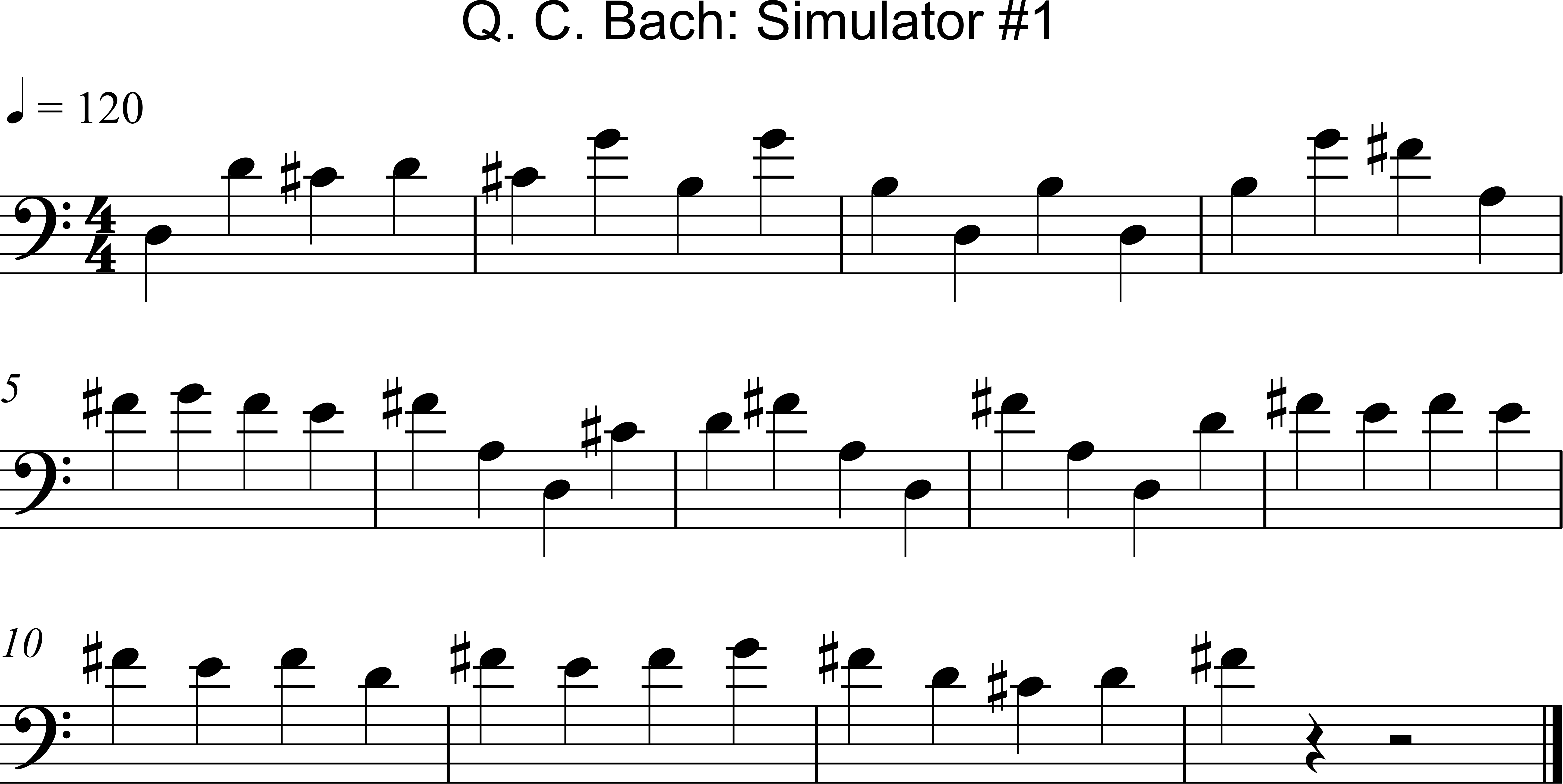}
\caption{Quantum generated Bach, simulator, $n = 1$.}
\label{fig:bach_simul1}
\end{center}
\end{figure}

Figures \ref{fig:bach_hard1} and \ref{fig:bach_simul1} shows the results from hardware and simulation, with $n=1$.

\begin{figure}[H]
\begin{center}\vspace{0.8cm}
\includegraphics[width=0.6\linewidth]{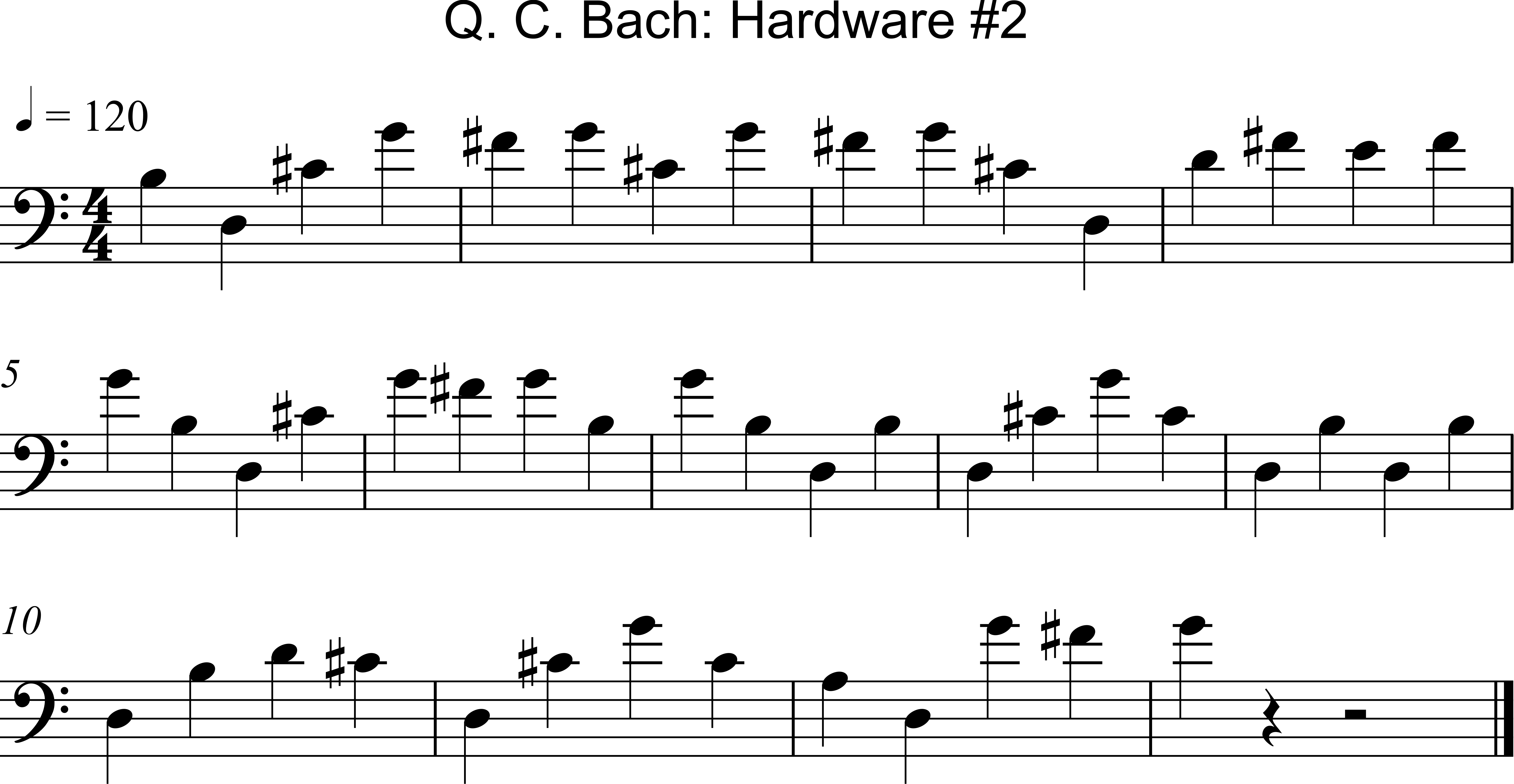}
\caption{Quantum generated Bach, hardware, $n = 2$.}
\label{fig:bach_hard2}
\end{center}
\end{figure}

\begin{figure}[H]
\begin{center}\vspace{0.8cm}
\includegraphics[width=0.6\linewidth]{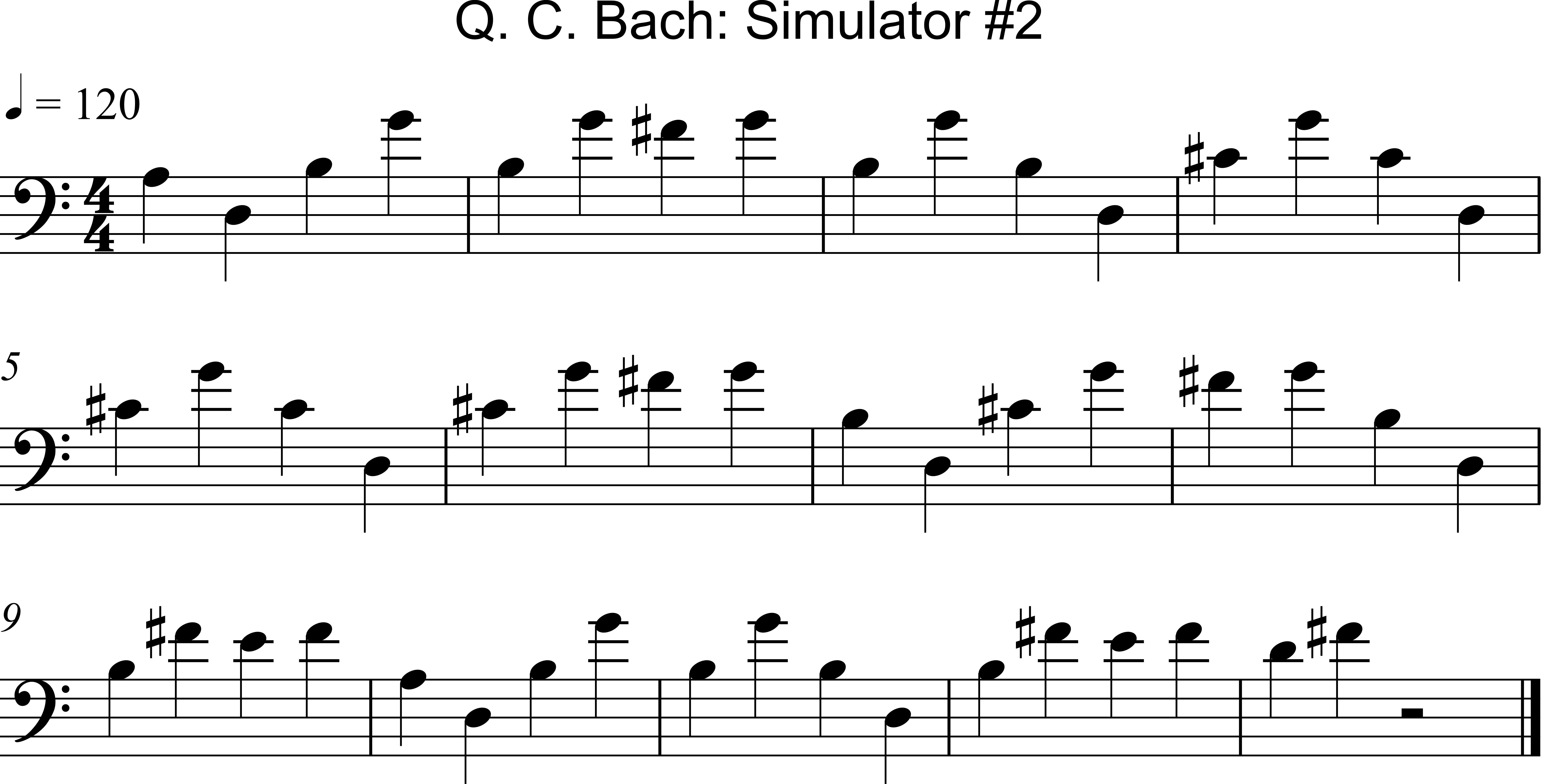}
\caption{Quantum generated Bach, simulator, $n = 2$.}
\label{fig:bach_simul2}
\end{center}
\end{figure}

Figures \ref{fig:bach_hard2} and \ref{fig:bach_simul2} shows the results from hardware and simulation, with $n=2$.

\begin{figure}[H]
\begin{center}\vspace{0.8cm}
\includegraphics[width=0.6\linewidth]{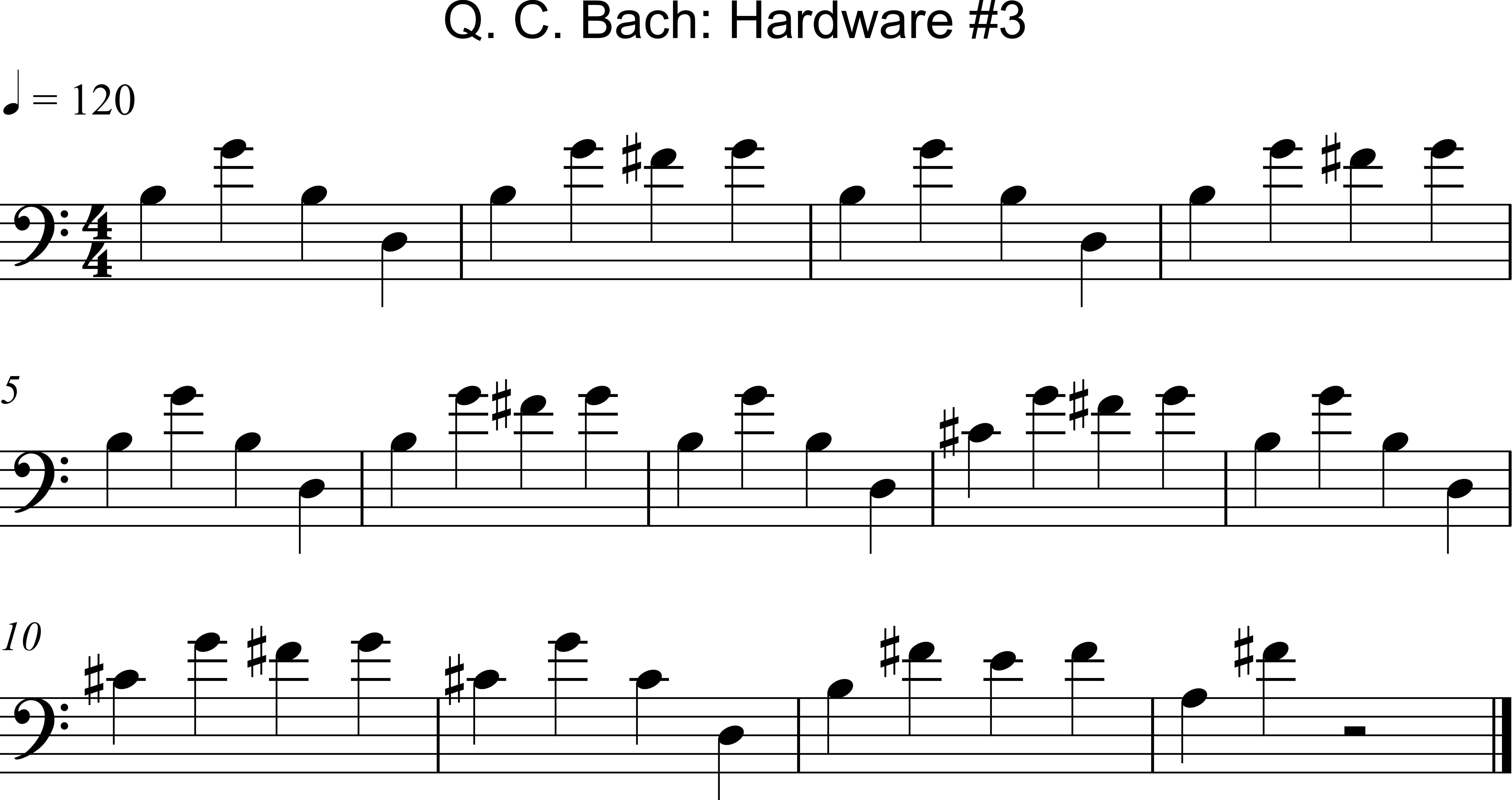}
\caption{Quantum generated Bach, hardware, $n = 3$.}
\label{fig:bach_hard3}
\end{center}
\end{figure}

\begin{figure}[H]
\begin{center}\vspace{0.8cm}
\includegraphics[width=0.7\linewidth]{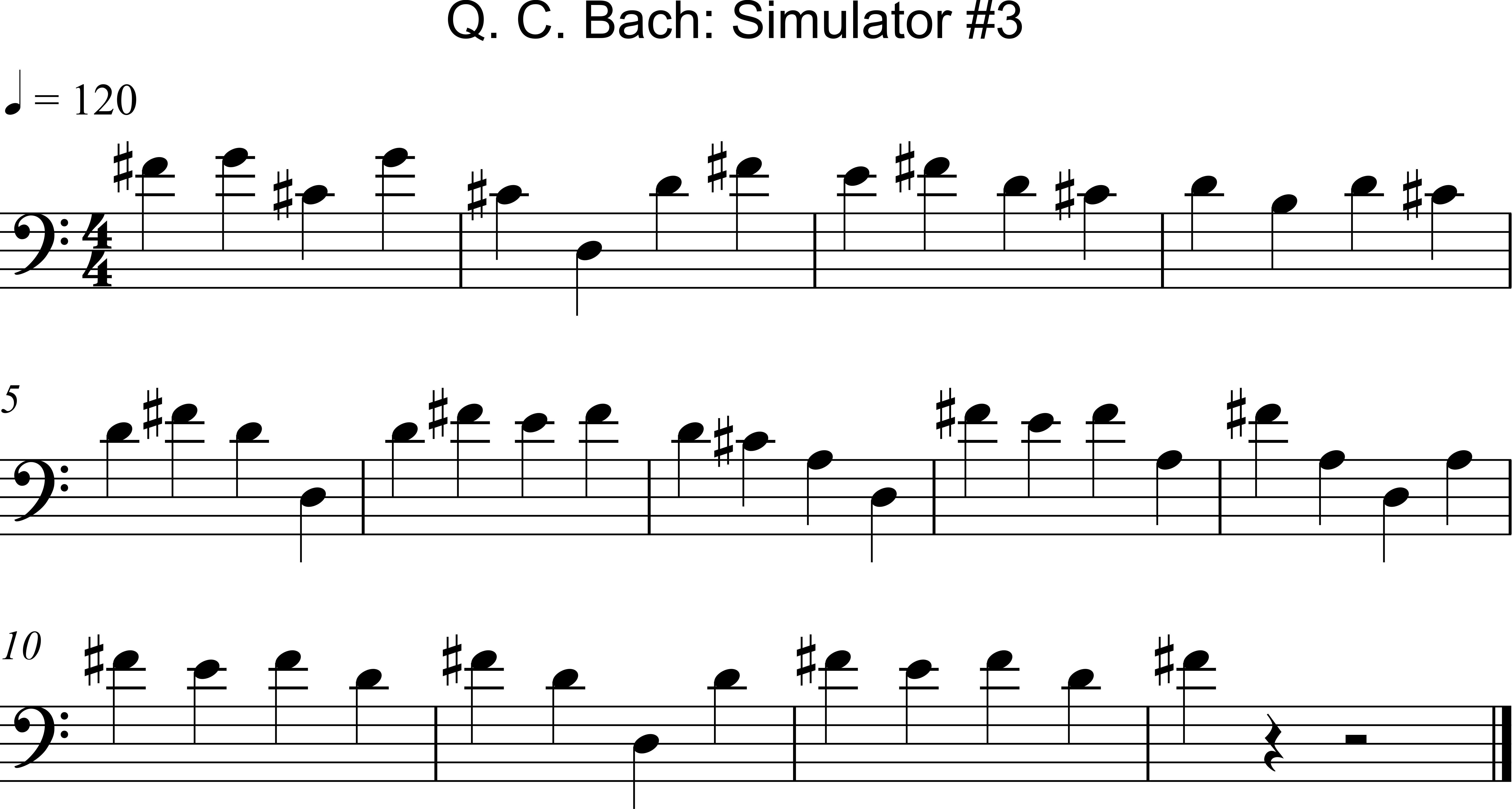}
\caption{Quantum generated Bach, simulator, $n = 3$.}
\label{fig:bach_simul3}
\end{center}
\end{figure}

Figures \ref{fig:bach_hard3} and \ref{fig:bach_simul3} shows the results from hardware and simulation, with $n=3$.

\newpage

%
%



\begin{thebibliography}{100}

\bibitem{Anumanchipalli2010}
Anumanchipalli, G., Cheng, Y., Fernandez, J., Huang, X., Mao, Q., and Black A. (2010).
\enquote{KLATTSTAT: Knowledge-based Parametric Speech Synthesis}.  
\textit{Proceedgins Seventh ISCA Tutorial and Research Workshop on Speech Synthesis} Kyoto, Japan.

\bibitem{Arya2022}
Arya, A., Botelho, L., Canete, F., Kapadia, D. and Salehi, O. (2022).
\enquote{Music Composition Using Quantum Annealing}.
In E. R. Miranda (Ed.), \textit{Quantum Computer Music: Foundations, Methods and Advanced Concepts}. Cham: Springer. 
Pre-print: \url{https://arxiv.org/abs/2201.10557} (Accessed 05 June 2022)

\bibitem{Bell2011}
Bell, C. (2011). \enquote{Algorithmic music composition using dynamic Markov chains and genetic algorithms}.
\textit{Journal of Computing Sciences in Colleges}, 27(2):99-107.

\bibitem{Bernhardt2019}
Bernhardt, C. (2019). \textit{Quantum Computing for Everyone}. 
The MIT Press. ISBN: 978-0262039253.

\bibitem{Bevins2013}
Bevins, G. (2013). 
\enquote{Computer technology in modern music: a study of current tools and how musicians use them}. 
\textit{Capstone Projects and Master's Theses}, 367. California State University. 
Digital Commons: \url{https://digitalcommons.csumb.edu/caps_thes/367/} (Accessed on 05 June 2022)

\bibitem{Bittner2022}
Bittner, R. (2022). \enquote{Meet Basic Pitch: Spotify’s Open Source Audio-to-MIDI Converter}.
\textit{Spotify Engineering Blog}.
Online publication: \url{https://engineering.atspotify.com/2022/06/meet-basic-pitch/} (Accessed on 05 June 2022).

\bibitem{Brown_etal_2015}
Brown, A. R., Gifford, T. and Davidson, R. (2015). \enquote{Techniques for Generative Melodies Inspired by Music Cognition}.
\textit{Computer Music Journal}, 39(1):11-26.

\bibitem{Brown1999}
Brown, S., Merker, B. and Wallin, N. L. (Eds.) (1999). \textit{The Origins of Music}. 
The MIT Press, USA. ISBN: 978-0262731430.

\bibitem{Chuharski2022}
Chuharski, J. M. (2022). \enquote{Adiabatic Quantum Computing and Applications to Music}.
In E. R. Miranda (Ed.), \textit{Quantum Computer Music: Foundations, Methods and Advanced Concepts}. Cham: Springer. 

\bibitem{Crosson2021}
Crosson, E. J. and Lidar, D. A. (2021).
\enquote{Prospects for quantum enhancement with diabatic quantum annealing}.
\textit{Nature Reviews Physics}, 3(466-489). 
DOI: https://doi.org/10.1038/s42254-021-00313-6

\bibitem{DodgeJerse1997}
Dodge, C. and Jerse, T. (1997). \textit{Computer Music: Synthesis, Composition, and Performance}.
Schirmer Books, 2rd edition. ISBN:978-0028646824.

\bibitem{Doornbusch2004}
Doornbusch, P. (2004). \enquote{Computer Sound Synthesis in 1951: The Music of CSIRAC.}
\textit{Computer Music Journal}, 28:1(10-25).

\bibitem{Grant2020}
Grant E. K. and Humble, T. S. (2020).
\enquote{Adiabatic Quantum Computing and Quantum Annealing}, 
\textit{Physics}, July 2020. 
Open access: \url{https://doi.org/10.1093/acrefore/9780190871994.013.32} (Accessed on 05 June 2022)

\bibitem{Griffiths2018}
Griffiths, D. J. and  Schroeter, D. F. (2018). \textit{Introduction to Quantum Mechanics}.
Cambridge University Press. ISBN: 9781107189638.

\bibitem{Grover1997}
Grover, L. K. (1997). \enquote{Quantum Mechanics Helps in Searching for a Needle in a Haystack}.
\textit{Physical Review Letters}, 79(2):325-328.
DOI: 10.1103/PhysRevLett.79.325

\bibitem{Grumbling2019}
Grumbling, E. and Horowitz, M (Eds.) (2019). Quantum Computing:
Progress and Prospects. National Academies Press. ISBN: 9780309479691.
DOI: https://doi.org/10.17226/25196

\bibitem{Hadimlioglu2018}
Hadimlioglu, I. A. and King, S. A. (2018). \enquote{Automated musical transitions through rule-based synthesis using musical properties}.
\textit{Entertainment Computing}, 28:59-67.

\bibitem{Hahn2020}
Hahn, M. (2020). \enquote{Subtractive Synthesis: Learn Synthesizer Sound Design}. \textit{LANDR Blog}. 
Available online: \url{https://blog.landr.com/subtractive-synthesis/} (Accessed on 05 June 2022).

\bibitem{Hamido2020}
Hamido, O.C., Cirilo, G. A. and Giusto, E. (2020).
\enquote{Quantum synth: a quantum-computing-based synthesizer}. 
\textit{Proceedings of the 15th International Conference on Audio Mostly}, Graz, Austria.
Open access: \url{https://dl.acm.org/doi/10.1145/3411109.3411135}

\bibitem{HillerIsaacson1959}
Hiller, L. A. and Isaacson, L. M. (1959). \textit{Experimental Music: Composition with an Electronic Computer}. 
McGraw-Hill. 
Available online: \url{https://archive.org/details/experimentalmusi00hill/page/n5/mode/2up} (Accessed on 07 May 2021)

\bibitem{IBMQuantum}
\url{https://www.ibm.com/quantum} (Accessed on 25 June 2022)

\bibitem{Kenmochi2007}
Kenmochi, H. and Ohshita, H. (2007).
\enquote{VOCALOID - Commercial singing synthesizer based on sample concatenation}.
\textit{Proceedings of INTERSPEECH 2007, 8th Annual Conference of the International Speech Communication Association}, Antwerp, Belgium.  

\bibitem{Kirke2017}
Kirke, A. and Miranda, E. R. (2017). 
\enquote{Experiments in Sound and Music Quantum Computing}. In E. R. Miranda (Ed.), \textit{Guide to Unconventional Computing for Music}. Cham, Switzerland: Springer, pp. 121-158. ISBN: 978-3319498805.

\bibitem{Klatt1980}
Klatt, D. H. (1980).
\enquote{Software for a cascade/parallel formant synthesizer}.
\textit{Journal of Acoustic Society of America}, 67(3):971-995.

\bibitem{M3}
\url{https://github.com/Qiskit-Partners/mthree} (Accessed on 25 June 2022)

\bibitem{Maestre2009}
Maestre, E., Ramirez, R., Kersten, S. and Serra, X. (2009).
\enquote{Expressive Concatenative Synthesis by Reusing Samples from Real Performance Recordings}.
\textit{Computer Music Journal}, 33(4):23-42.

\bibitem{Manning2004}
Manning, P. (2004).
\textit{Electronic and Computer Music}.
Oxford, UK: Oxford University Press. ISBN: 978-0195170856.

\bibitem{Mannone2022}
Mannone, M. and Rocchesso, D. (2022). 
\enquote{Quanta in Sound, the Sound of Quanta: A Voice-Informed Quantum Theoretical Perspective on Sound}. In: Miranda, E.R. (Ed.),  \textit{Quantum Computing in the Arts and Humanities}. Cham, Switzerland: Springer, pp. 193-226. ISBN: 978-3030955373.

\bibitem{mapomatic}
\url{https://github.com/Qiskit-Partners/mapomatic} (Accessed on 25 June 2022).

\bibitem{Mathews1970}
Mathews, M. V. and Moore, F. R.  (1970). \enquote{GROOVE—a program to compose, store, and edit functions of time}. 
\textit{Communications of the ACM}, 13(12):715-721. 
Open Access: \url{https://dl.acm.org/doi/10.1145/362814.362817} (Accessed on 25 June 2022)

\bibitem{McAdams1987}
McAdams, A. (ed.) (1987). Music and psychology: a mutual regard. \textit{Contemporary Music Review}, Vol. 2, Part 1.
Gordon and Breach, UK.

\bibitem{Miranda2022}
Miranda, E. R. and Basak, S. T. (2022). \enquote{Quantum Computer Music: Foundations and Initial Experiments}.
In E. R. Miranda (Ed.), \textit{Quantum Computer Music: Foundations, Methods and Advanced Concepts}. Cham: Springer. 
Pre-print: \url{https://arxiv.org/abs/2110.12408} (Accessed on 25 June 2022)

\bibitem{Miranda2021}
Miranda, E. R. (2021). \enquote{Quantum Computer: Hello, Music!}. 
In E. R. Miranda (Ed.). \textit{Handbook of Artificial Intelligence for Music: Foundations, Advanced Approaches, and Developments for Creativity.}
Springer International Publishing. ISBN: 9783030721152.
Pre-print: \url{https://arxiv.org/abs/2006.13849} (Accessed on 25 June 2022)

\bibitem{Miranda2021b}
Miranda, E. R. (Ed.) (2021). 
\textit{Handbook of Artificial Intelligence for Music: Foundations, Advanced Approaches, and Developments for Creativity.}
Springer International Publishing. 
ISBN: 9783030721152.

\bibitem{Miranda2020a}
Miranda, E. R. (2020). \enquote{Creative Quantum Computing: Inverse FFT, Sound Synthesis, Adaptive Sequencing and Musical Composition}. 
In A. Adamatzky (Ed.), \textit{Handbook of Unconventional Computing}, pp.493-523.
World Scientific. ISBN: 9789811235030.
Pre-print: \url{https://arxiv.org/abs/2005.05832} (Accessed on 25 June 2022)

\bibitem{Miranda2002}
Miranda, E. R. (2002). \textit{Computer Sound Design: Synthesis techniques and programming}. 
Elsevier Focal Press. ISBN: 978-0240516936.

\bibitem{Miranda2001}
Miranda, E. R. (2001). \textit{Composing Music with Computers}. 
Elsevier Focal Press. ISBN: 9780240515670.

\bibitem{MusicExamples}
SoundClick album: \url{https://www.soundclick.com/artist/default.cfm?bandID=1504503} (Accessed on 25 June 2022)

\bibitem{Nierhaus2009}
Nierhaus, G. (2009). \textit{Algorithmic Composition: Paradigms of Automated Music Generation}
Springer. ISBN: 978-3211755402.

\bibitem{QuantinuumTKET}
\url{https://github.com/CQCL/tket} (Accessed on 25 June 2022)

\bibitem{QiskitPassManager}
\url{https://qiskit.org/documentation/tutorials/circuits_advanced/04_transpiler_passes_and_passmanager.html} (Accessed on 25 June 2022)

\bibitem{QuSingQuTune}
QuSing GitHub repository: \url{https://github.com/iccmr-quantum/QuSing} (Accessed on 25 June 2022)

\bibitem{FFT}
Redman, N. (2002). \enquote{A gentle introduction to the FFT}. \textit{Ear Level Engineering}.
Online publication: \url{https://www.earlevel.com/main/2002/08/31/a-gentle-introduction-to-the-fft/} (Accessed on 01 July 2022)

\bibitem{Rieffel2011}
Rieffel, E. and Polak, W. (2011). \textit{Quantum Computing: A Gentle Introduction}.
The MIT Press. ISBN: 9780262015066.

\bibitem{Rutledge1995}
Rutledge, J. C., Cummings, K. E., Lambert, D. A.  and Clements, M. A. (1995). 
\enquote{Synthesizing styled speech using the Klatt synthesizer}. 
\textit{Proceedings of 1995 International Conference on Acoustics, Speech, and Signal Processing}. 
DOI: 10.1109/ICASSP.1995.479681.

\bibitem{Serra1990}
Serra, X. and Smith III, J. (1990).
\enquote{Spectral Modeling Synthesis: A Sound Analysis/Synthesis System Based on a Deterministic Plus Stochastic Decomposition}.
\textit{Computer Music Journal}, 14(4):12-24.

\bibitem{Siegelwax2021}
Siegelwax, B. N. (2021). \textit{Dungeons \& Qubits: An Adventurer's Tale Beyond the Quantum Computing Tutorials}.
Distributed on-line: \url{https://leanpub.com/dungeons-n-qubits} (Accessed on 25 June 2022)

\bibitem{Schwarz2005}
Schwarz. D. (2005). 
\enquote{Current Research in Concatenative Sound Synthesis}.
\textit{Proceedings of International Computer Music Conference (ICMC 2005)}, Barcelona, Spain. 
Available online: \url{https://hal.archives-ouvertes.fr/hal-01161337/file/index.pdf} (Accessed on 05 June 2022)

\bibitem{Shapiro2021}
Shapiro, I. and Huber, M. (2021). \enquote{Markov Chains for Computer Music Generation}.  
\textit{Journal of Humanistic Mathematics}, 11(2):167-195. 
DOI: 10.5642/jhummath.202102.08

\bibitem{Strategist2020}
Strategist, Q. (2020).
\enquote{What is Quantum Annealing and how does it differ from Gate based Quantum Computers?}.
\textit{Quantum Zeitgeist}, October 2020. Available online: \url{https://quantumzeitgeist.com} (Accessed on 08 June 2022).

\bibitem{Villavicencio2010}
Villavicencio, F. and Bonada, J. (2010).
\enquote{Applying voice conversion to concatenative singing-voice synthesis}.
\textit{Proceedings of 11th Annual Conference of the International Speech Communication Association}, Makuhari, Chiba, Japan.

\bibitem{Weaver2018}
Weaver, J. (2018). 
\enquote{Jamming with a Quantum Computer}. 
\textit{Rigetti Tech Blog}. 
Available online: \url{https://www.medium.com/rigetti/jamming-with-a-quantum-computer-bed05550a0e8} (Accessed on 25 June 2022)

\bibitem{Wikstrom2013}
Wikstrom, P. (2013). \enquote{The Music Industry in an Age of Digital Distribution}.
\textit{OpenMind BBVA}. Available online: \url{https://www.bbvaopenmind.com/en/articles/the-music-industry-in-an-age-of-digital-distribution/} (Accessed on 25 June 2022)

\bibitem{Wilson2017}
Wilson, M., Chandna, P., Daido, R. and Hisaminato, Y. (2017).
\enquote{Experiments in Making VOCALOID Synthesis Mode Human-like Using Deep Learning}.
\textit{IPSJ SIG Technical Reports}, Vol. 2017-MUS-114, No. 4.

\bibitem{Wong2022}
Wong, T. G. (2022). \textit{Introduction to Classical and Quantum Computing}. 
Rooted Grove. ISBN: 979-8985593105. 

\end{thebibliography}
\end{document}